\documentstyle[12pt]{article}
\begin{document}
\newcommand{\pe}{p_{e}}
\newcommand{\pxi}{p_{\tilde \chi_{i}}}
\newcommand{\pxj}{p_{\tilde \chi_{j}}}
\newcommand{\pl}{p_{\tilde l}}
\newcommand{\pqi}{p_{q_{in}}}
\newcommand{\pqo}{p_{q_{out}}}
\newcommand{\mxi}{m_{\tilde \chi_{i}}}
\newcommand{\mxj}{m_{\tilde \chi_{j}}}
\newcommand{\ml}{m_{\tilde l}}
\newcommand{\mq}{m_{\tilde q}}
\hfill
UWITP-94/7  \\
\hspace*{\fill}
UH-511-815-94   \\
\vspace{0.5cm}
\begin{center}
{\bf {\LARGE Associated Slepton-Neutralino/Chargino Production
at LEP$\otimes$LHC}} \\[0.7cm]
{\bf T. W\"ohrmann\footnote{email:
THOMAS@uhhepg.phys.hawaii.edu, Telefax: (808) 956-2930} }\\
{\em Institut f\"ur Theoretische Physik,
Universit\"at W\"urzburg,
Am Hubland,
97074 W\"urzburg, Germany}\\and \\
{\em University of Hawaii, Department of Physics and Astronomy,
Watanabe Hall, 2505 Correa Road, Honolulu, Hawaii 96822, USA }\\
{\bf H.Fraas}\\ {\em Institut f\"ur Theoretische Physik,
Universit\"at W\"urzburg,
Am Hubland,
97074 W\"urzburg, Germany}\\[0.7cm]
{\bf {\large Abstract }}
\end{center}
\hspace{0.5cm}
We examine for representative gaugino-higgsino mixing scenarios
slepton-neutralino and slepton-chargino production in deep
inelastic ep-scattering at $\sqrt{s}=1.8$ TeV. We find
sneutrino-chargino production to be the dominant process with cross
sections more than one order of magnitude bigger than those for
slepton-squark production. Also for associated production of
sneutrinos and zino-like neutralinos the cross sections are at least
comparable to those for $\tilde l \tilde q$-production, whereas
selectron-neutralino/chargino production is with cross sections
significantly smaller than those for selectron-squark production less
favorable. Typical signatures include events with up to four charged leptons,
hadronic jets and, in some cases, gauge bosons.
\vfill
\begin{center}
December 1994
\end{center}
\newpage
\noindent
\section{ Introduction }\label{int}
Supersymmetry is considered to be one of the most attractive extensions
for physics beyond the Standard Model. Searching for supersymmetric
particles will therefore play an important role also in the experimental
program of the future $ep$-collider LEP$\otimes $LHC. It especially
provides a very good possibility to search for the scalar partners of
electrons and neutrinos: Since in the simplest phenomenological model
which implements the idea of supersymmetry, the {\em Minimal
Supersymmetric Standard Model (MSSM)}, $R$-parity is conserved a
selectron or a sneutrino is always produced with a second SUSY-particle,
which can be either a squark or a neutralino or chargino respectively.
If the sum of the masses of the slepton and squark is smaller than 600
GeV, then $ep\rightarrow \tilde e \tilde q X $ and $ep\rightarrow \tilde
\nu \tilde q X$ are the most promising processes to search for
SUSY-particles at LEP$\otimes $LHC \cite{bartl5}. If, however, squarks
are heavy but sleptons are relatively light, than these processes are
suppressed or even inaccessible and the associated production of a
slepton and a neutralino $\tilde \chi_{i}^{0}\, (i=1,\ldots,4)$ or a
slepton and a chargino $\tilde \chi_{i}^{\pm}\, (i=1,2)$, which lead to new
interesting signatures, become the most important processes at
$ep$-colliders \cite{bartl6}.

At the parton level the possible production channels are
(a) $ eq \rightarrow \tilde e \tilde \chi_{i}^{0} q $,
(b) $ eq \rightarrow \tilde e \tilde \chi_{i}^{+} q' $,
(c) $ eq \rightarrow \tilde e \tilde \chi_{i}^{-} q' $,
(d) $ eq \rightarrow \tilde \nu \tilde \chi_{i}^{0} q' $,
(e) $ eq \rightarrow \tilde \nu \tilde \chi_{i}^{-} q $. Expecting that
the production of the lightest supersymmetric particle (LSP) $\tilde
\chi_{1}^{0}$ in processes (a) and (d) is particularly favorable, only
the associated production of the LSP has been analyzed in detail in the
deep inelastic region assuming, further, that the LSP is photino-like
\cite{ammr}. The couplings of the neutralinos and charginos, however,
sensitively depend on their nature which is determined by the way gauginos and
higgsinos are mixed. Therefore the question which of the four neutralinos
or two charginos will be produced with the highest rate sensitively
depends on the mixing scenario. Further the chargino production in the
process (e) can proceed via photon and $Z^0$ exchange, which can lead to
detectable cross sections also for associate chargino production.

We therefore investigate all five production channels and show for
LEP$\otimes $LHC energy ($\sqrt{s}=1.8$ TeV) the total cross sections for
three representative gaugino-higgsino mixing scenarios and for different
ratios of slepton and squark masses. Our choice of scalar masses is
partly motivated by renormalization
group relations coupling the sfermion masses and the gaugino mass
parameter of the MSSM. It turns out that the cross sections especially
for $ep\rightarrow \tilde \nu \tilde \chi_{i}^{-} X$ may attain values
between one or two orders of magnitude bigger than those for
slepton-squark production.

Further, the associated production of squarks and neutralinos/charginos
could be of interest, especially in case of the same mass for all sfermions.
Because squarkproduction, however, leads in many scenarios to signatures
with many hadronic jets, associated squark-neutralin/chargino production
is less favorable and not subjekt of this paper.

In order to see if these reactions might be suitable for providing us
with a SUSY signal at LEP$\otimes $LHC we include some remarks on the
competing Standard Model (SM) backgrounds and a brief discussion of the decay
patterns of the particles produced and the ensuing signatures for the
five reaction channels. Angular distributions and energy spectra would
be appropriate observables to extract SUSY events from SM backgrounds.
We therefore give in the appendix a complete list of the amplitudes
squared which also enables us to extend our investigations to polarized
electron beams.
\section{The Minimal Supersymmetric Standard Model (MSSM)}\label{MSSM}
We briefly describe those parts of the MSSM, which will be used for
our calculations in section \ref{analytic}. We will use the notation of
\cite{HaKa}. From this we get for the neutralino mass term
\begin{equation}
{\cal L}^{m}_{\chi^0} =-\frac{1}{2} m_Z (\psi_{i}^{0})^T
Y_{ij} \psi^{0}_{j} +h.c..
\end{equation}
With the spinors $\psi_{i}^{0}$ of the photino, the zino and the neutral
higgsinos,
\begin{equation}\label{spinor}
\psi_{i}^{0}=(-i\lambda_{\gamma },-i\lambda_Z ,
\psi^{1}_{H_1 }\cos \beta - \psi^{2}_{H_2 }\sin \beta  ,
\psi^{1}_{H_1 }\sin \beta + \psi^{2}_{H_2 }\cos \beta),
\quad i=1,\ldots,4,
\end{equation}
and the mass matrix $Y_{ij}$
\begin{equation}
{\bf Y}=\left( \begin{array}{cccc}
M (\sin^2 \theta_W +\frac{M'}{M} \cos^2 \theta_W )&
M (1-\frac{M'}{M} ) \sin \theta_W \cos \theta_W & 0 & 0 \\
M(1-\frac{M'}{M} ) \sin \theta_W \cos \theta_W &
M (\cos^2 \theta_W +\frac{M'}{M} \sin^2 \theta_W )& m_Z & 0 \\
0 & m_Z & \mu \sin 2\beta &-\mu \cos 2\beta \\
0 & 0 &-\mu \cos 2\beta & -\mu \sin 2\beta \end{array} \right)
\end{equation}
We may diagonalize this matrix with the real symmetric matrix $N_{ij}$,
where the columns of this matrix are given by the neutralino mass
eigenstates $ \chi_{i}^{0} $ in the basis of eq. (\ref{spinor})\cite{bartl1}:
\begin{equation}
\chi_{i}^{0}=N_{ij}\psi_{j}^{0},\quad N_{im}N_{jn}Y_{mn}=\eta_i m_i
\delta_{ij}.  \end{equation}
with $\eta_i$ a signfactor, which arise because the eigenvalues from
$Y_{ij}$ may be negative.
\begin{equation}\label{nffkop}
\chi_{i}^{0} = \frac{1}{N_i} \left(\begin{array}{c}
\frac{M}{m_Z}(1-\frac{M'}{M}) \sin \theta_W \cos \theta_W
(m_{i}^{2} -\mu^2) \\ \frac{1}{m_Z}
\left( m_i -M (\sin^2 \theta_W +\frac{M'}{M} \cos^2 \theta_W )\right)
(m_{i}^{2} -\mu^2 ) \\
\left( m_i -M (\sin^2 \theta_W +\frac{M'}{M} \cos^2 \theta_W )\right)
(m_i +\mu \sin 2\beta )\\
\left( m_i -M (\sin^2 \theta_W +\frac{M'}{M} \cos^2 \theta_W )\right)
(-\mu \cos 2\beta) \end{array} \right),
\end{equation}
with
\begin{eqnarray*}
N_i&=& \left( \frac{(m_{i}^{2} -\mu^2)^2}{m_{Z}^{2}}
\left(\sin^2 \theta_W (m_i -M )^2
+\cos^2 \theta_W (m_i -M' )^2 +(m_i -M )^2
(m_i -M' )^2 \right) \right. \\
& & +\left( m_i -M (\sin^2 \theta_W +\frac{M'}{M} \cos^2 \theta_W )
\right)^2(\mu \sin^2 \beta )^2 \left. \right)^{1/2} .
\end{eqnarray*}

The chargino mass sector is described by
\begin{equation}
{\cal L}^{m}_{ \chi^{\pm }}=-\frac{1}{2}\left( \psi^+ \psi^- \right)
\left( \begin{array}{cc} 0 & X^T \\ X & 0 \end{array} \right)
\left(\begin{array}{c}  \psi^+ \\  \psi^- \end{array}
\right) +h.c., \end{equation}
with the mass matrix
\begin{equation}
X=\left( \begin{array}{cc} M & m_W \sqrt{2} \sin \beta \\ m_W \sqrt{2}
\cos \beta & \mu \end{array} \right)
\end{equation}
and the two-component spinors $\psi^{\pm }_{j}$ of the winos and charged
higgsinos
\begin{equation}
\psi^{+}_{j}=(-i\lambda^+ , \psi^{1}_{H_2} ),\qquad
\psi^{-}_{j}=(-i\lambda^- , \psi^{2}_{H_1} ),\qquad j=1,2.
\end{equation}
We may diagonalize $X$ by unitary $2\times 2$-matrices $U$ and $V$
\begin{equation}
U^{\ast }_{im} V_{jn} X_{mn} = \eta_i m_i \delta_{ij},
\end{equation}
with the mass eigenstates
\begin{equation}
\chi^{+}_{i} =V_{ij} \psi^{+}_{j},\qquad
\chi^{-}_{i} =U_{ij}\bar { \psi^{-}_{j}} .
\end{equation}
Notice that like in the neutralino case we get also
positive or negative mass eigenvalues with \cite{bartl1}
\begin{equation}
\eta_{1,2}m_{1,2}=\frac{1}{2}\left( \sqrt{(M-\mu )^2
+2m_{W}^{2}(1+\sin 2\beta )} \mp
\sqrt{(M+\mu )^2 +2m_{W}^{2}(1-\sin 2\beta )} \right).
\end{equation}
The matrix elements $U_{ij}$ and $V_{ij}$ are
\begin{eqnarray}\label{cffkop}
U_{12}&=&U_{21}=\frac{\theta_1 }{\sqrt{2}} \sqrt{1+\frac{M^2-\mu^2 -
2m_{W}^{2}\cos 2\beta }{W} } ,\\
U_{22}&=&-U_{11}=\frac{\theta_2 }{\sqrt{2}} \sqrt{1-\frac{M^2-\mu^2 -
2m_{W}^{2}\cos 2\beta }{W} } ,\\
V_{21}&=&-V_{12}=\frac{\theta_3 }{\sqrt{2}} \sqrt{1+\frac{M^2-\mu^2 +
2m_{W}^{2}\cos 2\beta }{W} } ,\\
V_{11}&=&V_{22}=\frac{\theta_4 }{\sqrt{2}} \sqrt{1-\frac{M^2-\mu^2 +
2m_{W}^{2}\cos 2\beta }{W} } ,
\end{eqnarray}
with $\qquad W=\sqrt{(M^2+\mu^2 +2m_{W}^{2} )^2 -4(M \mu -m_{W}^{2}
\sin 2\beta )^2 }$. \\
The sign factors $\theta_i, i=1,\ldots,4$ are given in table 1.\\
For further details of this mixings see e.g. \cite{bartl1}.

The couplings in the Feynman graphs follow from the lagrangian of the
MSSM. The sfermion-fermion-neutralino couplings we get from
\begin{equation}
{\cal L}_{f \tilde f \tilde \chi^{0} }=\sum_{ij} \frac{1}{i}\left(
(\eta^{\ast 0}_{f_{iL}} )_j \bar f_{iR} \tilde \chi^{0}_{j} \tilde f_{iL}+
(\eta^{\ast 0}_{f_{iR}} )_j \bar f_{iL} \tilde \chi^{0}_{j} \tilde f_{iR}
\right) +h.c.,
\end{equation}
with $f_{iR,L}=P_{R,L} f_i,\, P_{R,L}=\frac{1}{2} (1\pm \gamma_5 )$ and
\begin{eqnarray} \label{KopNeu}
\left(\eta^{0}_{f_{iL}}\right)_{j}&=&-ig\sqrt{2}\left(\frac{1}{\cos\theta_{W}}
\left(T_{3f_{i}}-e_{f_{i}}\sin^{2}\theta_{W}\right)N_{j2}+
e_{f_{i}}\sin\theta_{W}N_{j1}
\right)\nonumber \\
\left(\eta^{0}_{f_{Ri}}\right)_{j}&=&-ig\sqrt{2}e_{f_{i}}
\sin\theta_{W}\left(\tan
\theta_{W}N^{\ast}_{j2}-N^{\ast}_{j1}\right) \qquad j=1,\ldots ,4.
\end{eqnarray}
With $g=e/\sin \theta_W,\,e>0$ and $e_f,\,T_{3f}$ the charge and the
third component of the weak isospin of the fermion $f$.
The fermion-sfermion-chargino couplings we get from
\begin{equation}
{\cal L}_{f \tilde f \tilde \chi^{\pm } }=-g\sum_i \left(
\bar u_L U_{i1}\tilde \chi_i \tilde d_L +
\bar d_L V_{i1}\tilde \chi_{i}^{c} \tilde u_L \right)+h.c.
\end{equation}
where we also introduce the couplings $(\eta_{f})_i$,
\begin{eqnarray} \label{KopCh}
\left(\eta^{c}_{f_{R}}\right)_{i}&=&0 \nonumber \\
\left(\eta^{c}_{u_{L}}\right)_{i}&=&-igU_{i1} \nonumber \\
\left(\eta^{c}_{d_{L}}\right)_{i}&=&-igV^{\ast}_{i1}  \qquad
i=1,2,
\end{eqnarray}
where the $u$ ($d$) are the uptype (downtype) fermions.\\
The couplings of a gauge boson with two neutralinos or charginos we get
from
\begin{eqnarray}
{\cal L}_{W^{-} \tilde \chi^+ \tilde \chi^0 }&=&-iW_{\mu}^{-} \bar {\tilde
\chi_{i}^{0} } \gamma^{\mu }(O_{ij}^{L} P_L +O_{ij}^{R} P_R )\tilde
\chi^{+}_{j}\\
{\cal L}_{Z \tilde \chi^+ \tilde \chi^- }&=& -iZ_{\mu } \bar {\tilde
\chi_{i}^{+} } \gamma^{\mu} (O'^{L}_{ij} P_L + O'^{R}_{ij} P_R ) \tilde
\chi_{j}^{+}, \\
{\cal L}_{Z \tilde \chi^0 \tilde \chi^0 }&=&\frac{-i}{2} Z_{\mu }\bar
{\tilde \chi_{i}^{0} } \gamma^{\mu} (O''^{L}_{ij} P_L + O''^{R}_{ij} P_R )
\tilde \chi_{j}^{0}\\
{\cal L}_{\gamma \tilde \chi^+ \tilde \chi^- }&=&-e A_{\mu } \bar {\tilde
\chi_{i}^{+} } \gamma^{\mu } \tilde \chi_{i}^{+}.
\end{eqnarray}
with
\begin{eqnarray}\label{WCNKop}
O^{L}_{ij}&=&-\frac{ig}{\sqrt{2}}(\sin\theta_V N_{i4}
-\cos\theta_V N_{i3})V^{\ast}_{j2}+g(\sin\theta_W N_{i1}
+\cos\theta_W N_{i2})V^{\ast}_{j1}, \nonumber \\
O^{R}_{ij}&=&\frac{ig}{\sqrt{2}}(\cos\theta_V
N_{i4}^{\ast}+\sin\theta_V N_{i3}^{\ast})U_{j2}+
g(\sin\theta_W N_{i1}^{\ast}+\cos\theta_W N^{\ast}_{i2})U_{j1}, \\
O'^{L}_{ij}&=&\frac{ig}{\cos \theta_W}\left(
\delta_{ij}\sin^{2}\theta_W -V_{i1} V_{j1}^{\ast } -\frac{1}{2}
V_{i2} V_{j2}^{\ast}\right), \nonumber \\
O'^{R}_{ij}&=&\frac{ig}{\cos \theta_W}\left(
\delta_{ij}\sin^{2}\theta_W-U_{i1}^{\ast}U_{j1}-\frac{1}{2}
U_{i2}^{\ast}U_{j2}\right)\\
O''^{L}_{ij}&=&\frac{ig}{2\cos\theta_W} \left( (
N_{i3}N^{\ast}_{j3}-N_{i4}N^{\ast}_{j4})\cos 2\theta_V
-(N_{i3}N^{\ast}_{j4}
+N_{i4}N^{\ast}_{j3})\sin 2\theta_V \right),\nonumber \\
\label{GCNKop}
O''^{L}_{ij}&=&O''^{R\ast}_{ij}
\end{eqnarray}
Finally the sfermion-sfermion-gauge boson couplings we get from
\begin{eqnarray}
{\cal L}_{\tilde f \tilde f V}&=&\frac{-ig}{\sqrt{2}}\left(W_{\mu}^{+}
( \tilde u^{\ast}_{L} \stackrel{\leftrightarrow}{\partial^{\mu}} \tilde
d_L )+ (W_{\mu}^{-}( \tilde
d^{\ast}_{L}\stackrel{\leftrightarrow}{\partial^{\mu }}
\tilde u_L )\right) -\nonumber \\
& & \frac{ig}{\cos \theta_W} Z_{\mu} \sum_i (T_{3i}-e_i
\sin^{2} \theta_W )\tilde f^{\ast}_{i} \stackrel{\leftrightarrow}{
\partial^{\mu}} \tilde f_i
-ieA_{\mu }\sum_i e_i \tilde f_{i}^{\ast } \stackrel{\leftrightarrow}{
\partial^{\mu }} \tilde f_i.
\end{eqnarray}

In the MSSM also the left and right-handed sfermions mix. Because these
mixings are growing up with increasing fermion masses, these mixings are
only important for the top squarks.
We will also assume $R$-parity conservation, where $R$ is
given by $R=(-1)^{L+3B+2S}$ with $L$ the lepton number, $B$ the baryon
number and $S$ the spin of the particle.

Assuming a common gaugino mass term at the unification
scale, it follows
$M'/M=\alpha=5/3 \tan^2 \theta_W$. From the assumption of a unique soft
symmetry breaking term at the unification scale it follows for the
gluinomass $m_{\tilde g}= M\sin^2 \theta_W \alpha_s /\alpha_{em}$.
Assuming a common sfermion mass term at the unification scale (like this
is valid in supergravity models), it follows from \cite{Pol} a relation
connecting the sfermion with the gaugino/higgsino masses, which we will
use in some of our scenarios
\begin{equation}\label{eqpol}
m^{2}_{\tilde f_{L,R}}=m^{2}_{f} +m^{2}_{0}+C(\tilde f)M^2\pm m^{2}_{Z}
\cos 2 \beta )(T_{3f}-e_{f}\sin^2 \theta_W ),
\end{equation}
with $C(\tilde q_{R,L})\simeq
10$, $C(\tilde l_{R})\simeq 0.23$ and $C(\tilde l_{L})\simeq 0.79$.
\section{ Analytical Results }\label{analytic}
We will now consider the analytical calculations for the processes
$ep\rightarrow \tilde l \tilde \chi_{i}^{0,\pm} X $.
The Feynman graphs related to
the basic subprocesses $eq\rightarrow \tilde l \tilde \chi_{i}^{0,\pm} q $
are shown in fig. 1. The corresponding amplitudes are:
\begin{eqnarray} \label{eqeb1}
({\cal M}_{1})_{ab} & = & \bar u_{\tilde \chi_{i}}(p_{\tilde
\chi_{i}})\left(
\eta_{e_{a}}\right)_{i}u_{e_{a}}(p_{e})\bar u_{q_{out,b}}(p_{q_{out}})
\gamma_{\mu}u_{q_{in,b}}(p_{q_{in}})\left(p_{\tilde l}+p_{e}
-p_{\tilde \chi_{i}}\right)^{\mu} \nonumber\\
& & \sum_{X}\left( (\Delta_{X})(p_{q_{in}},
p_{q_{out}})\left(f_{e}\right)_{X }\left(f_{q}\right)_{X }\right)
\frac{i}{\left(p_{e}-p_{\tilde \chi_{i}}\right)^{2}-m_{\tilde l}^{2}}
 \\[0.2cm]
({\cal M}_{2})_{ab} & = & \sum_{j}\bar
v_{e_{a}}(p_{e})\left(\eta_{e_{a}}
\right)_{j}i\frac{\not \! p_{\tilde l}-\not \! p_{e}+m_{\tilde
\chi_{j}}}
{\left(p_{\tilde l}-p_{e}\right)^{2}-m^{2}_{\tilde \chi_{j}}}\left(
\eta_{q_{b}}\right)_{j}u_{q_{in,b}}(p_{q_{in}}) \nonumber\\
& & \frac{i}{\left(p_{q_{out}}+
p_{\tilde \chi_{i}}\right)^{2}-m^{2}_{\tilde q}+ i m_{\tilde q}
\Gamma_{\tilde q}^{tot}}\bar u_{q_{out,b}}
(p_{q_{out}})\left(\eta_{q_{b}}^{\ast}\right)_{i}v_{\tilde
\chi_{i}}(p_{\tilde \chi_{i}})\\[0.2cm]
({\cal M}_{3})_{ab} & = & \bar u_{\tilde \chi_{i}}(p_{\tilde \chi_{i}})
i\frac{\not \! p_{\tilde \chi_{i}}+\not \! p_{\tilde l}
+m_{l}}{\left(p_{\tilde
\chi_{i}}+p_{\tilde
l}\right)^{2}-m_{l}^{2}}\gamma_{\mu}u_{e_{a}}(p_{e})\left(
\eta_{e_{a}}\right)_{i}\bar u_{q_{out,b}}(p_{q_{out}})
\gamma^{\mu}u_{q_{in,b}}(p_{q_{in}}) \nonumber\\
& & \sum_{X}\left((\Delta_{X})
(p_{q_{in}},p_{q_{out}})\left(f_{e}\right)_{X }
\left(f_{q}\right)_{X }\right)\\[0.2cm]
({\cal M}_{4})_{ab} & = & -\sum_{j}\bar u_{q_{out,b}}(p_{q_{out}})
\left(\eta_{q_{b}}^{\ast}\right)_{j}
i\frac{\not \! p_{e}-\not \! p_{\tilde l}+m_{\tilde \chi_{j}}}{\left(
p_{e}-p_{\tilde l}\right)^{2}-m^{2}_{\tilde \chi_{j}}}\left(
\eta_{e_{a}}\right)_{j}u_{e_{a}}(p_{e})\nonumber \\
 & & \frac{i}{\left(p_{q_{in}}-
p_{\tilde \chi_{i}}\right)
^{2}-m^{2}_{\tilde q}}
\bar u_{\tilde \chi_{i}}(p_{\tilde \chi_{i}})
\left(\eta_{q_{b}}\right)_{i}u_{q_{in,b}}(p_{q_{in}}) \\[0.2cm]
\label{eqeb5}
({\cal M}_{5})_{ab} & = & \sum_{j}
\bar u_{\tilde \chi_{i}}(p_{\tilde \chi_{i}})\gamma_{\mu}
\sum_{X}\left((\Delta_{X})(p_{q_{in}},
p_{q_{out}})\left(f_{q}\right)_{X}\left({\cal
O}_{X}\right)_{ij}\right)\nonumber\\
& & i\frac{\not \! p_{e}-
\not \! p_{\tilde l}+m_{\tilde \chi_{j}}}{\left(p_{e}-
p_{\tilde l}\right)^{2}-m^{2}_{\tilde \chi_{j}}}
\left(\eta_{e_{a}}\right)_{i}u_{e_{a}}(p_{e})
\bar u_{q_{out,b}}(p_{q_{out}})\gamma^{\mu}u_{q_{in,b}}(p_{q_{in}})
\end{eqnarray}
The momenta of the incoming quark, outgoing quark, the electron, the
slepton and the neutralino/chargino are denoted by
$p_{q_{in}}$, $p_{q_{out}}$, $p_{e}$, $p_{\tilde l}$ and $p_{\tilde
\chi_{i}}$. The index $i$ denotes the outgoing neutralino/chargino and
the indices $a=R,L$ and $b=R,L$ denote the polarisation of the electron
and the incoming quark, respectively. Further
\[ \Delta_X(p_{q_{out}}-p_{q_{in}})
= \frac{-i}{(p_{q_{out}}-p_{q_{in}})^{2}-m_{X}^{2}} \]
Where $X=\gamma,\,Z^0,\,W$ denotes the gauge boson exchanged in the
graphs 1, 3 and 5. Because $X$ depends on the process, we show in table 2
this connection between exchanged gauge boson and process.

In eqs. (\ref{eqeb1})--(\ref{eqeb5}) $(f_f)_X$ are the electro weak
couplings of the fermion $f$ and the gauge boson $X$ with
\begin{eqnarray}\label{SUKop}
(f_f)_{\gamma}&=&-iee_f, \nonumber \\
(f_{f_{R}})_Z&=&ie\tan\theta_W  e_f, \nonumber \\
(f_{f_{L}})_Z&=&-i\frac{e}{\sin\theta_W \cos\theta_W}\left(T_{3f}
-\sin^{2}\theta_W e_f\right), \nonumber \\
(f_{f_{L}})_W&=&-i\frac{e}{\sqrt 2\sin \theta_W}.
\end{eqnarray}
The supersymmetric couplings from eqs. (\ref{KopNeu}) and (\ref{KopCh})
are denoted by $(\eta_f )_{i}$ and these from eqs.
(\ref{WCNKop})--(\ref{GCNKop}) are
denoted by $({\cal O}^{R,L}_{X})_{ij}$ with
\[
({\cal O}_X )_{ij}=({\cal O}^{R}_{X})_{ij}\frac{1+\gamma_5}{2}
+({\cal O}^{L}_{X})_{ij}\frac{1-\gamma_5}{2}.
\]
For selectron and sneutrino production we have in the amplitudes
${\cal M}_2$, ${\cal M}_4$ and ${\cal M}_5$ to sum the contributions
from the exchange of all four neutralinos and both charginos
respectively. Since the squark exchanged in graph 2 may approach its
mass shell in the accessible region of the phase space, the squark width
$\Gamma^{tot}_{\tilde q}$ enters in ${\cal M}_2$.

Depending on the production channel and the polarisation of the sleptons
and squarks certain amplitudes vanish. Thus for the process (b) and (d)
one obtains contributions from left-handed $u$-quarks and right-handed $\bar
d$-quarks only, whereas only left-handed $d$-quarks and right-handed
$\bar u$-quarks
are contributing to the process (c). To the process (e) all quarks
contribute to the terms ${\cal M}_{1,3,5}$ but only left-handed
$u$-quarks ($d$-quarks) and right-handed $\bar d$-quarks ($\bar u$-quarks) are
contributing to the amplitude ${\cal M}_2$ (${\cal M}_4$). In process
(a) all quarks contribute. Further for the processes (d) and (e) as well
as for process (b)/(c) and the amplitude ${\cal M}_3$/${\cal M}_1$ we
obtain contributions from left-handed electrons only. Finally the
amplitude ${\cal M}_1$ in case of process (b) and ${\cal M}_3$ in case
of process (c) vanish.

The amplitudes squared are completely listed in appendix A.
In order to obtain the total cross section, we average over the
polarisations of the incoming particles, sum over the polarisations of
the outgoing particles, fold this with the appropriate quark
distributions, sum over all partons of the proton and integrate over the
whole phase space:
\begin{eqnarray} \label{integ}
\sigma^{tot}\left(ep\rightarrow \tilde l \tilde \chi_i X\right) &=&
\sum_k \int \frac{ {\mbox d} x
}{8(E_{cm}^{eq})^2 (2\pi )^{5}}\frac{1}{4}\sum_{a,b}
\int \frac{{\mbox d}^3 \pl}{2E_{\tilde l}}\int \frac{{\mbox d}^3
\pxi}{2E_{\tilde \chi_i}}\int \frac{{\mbox d}^3 \pqo}{2E_{q_{out}}}
\times \nonumber \\
 & & q_k(x,Q^2)\left| \sum_n ({\cal M}_n)_{ab} \right|^2
\delta^4(\pe +\pqi -\pl -\pxi -\pqo )
\end{eqnarray}
The specific parametrization of the momenta in the phase space is, as well
as the limits of the phase space integration, given in appendix B.
\section{Numerical Results}\label{num}
To show the importance of gaugino-higgsino mixing
we shall present numerical results for the total production cross section
at $\sqrt{s} =1.8$ TeV, for
three representative gaugino higgsino mixing scenarios, shown in
table 3. For the SUSY parameters we assumed
$M'/M=\frac{3}{5}\tan\theta_W$, $m_{\tilde g}=M\sin^2 \theta_W
\alpha_s/\alpha_{em}\simeq 3M$ with $\sin^2 \theta_W=0.228$ and
$\alpha_s=0.1$, and $\tan \beta =v_2 /v_1 =2$  (notice
that the numerical results are not very sensitive to the value of
$\tan \beta $). For the masses of the gauge bosons we used $m_Z=91.2$
GeV and $m_W=80.1$ GeV.

The crucial difference lies in the nature of the neutralino states and
the chargino states. In scenario (A) the lightest neutralino $\tilde
\chi_{1}^{0}$ is almost a photino, whereas in scenario (C) it is nearly
a higgsino and in scenario (B) it is a zino-photino-higgsino mixture.
Similarly the light (heavy) chargino is wino-like (higgsino-like) in
scenario (A), higgsino-like (wino-like) in scenario (C), whereas in
scenario (B) both charginos are wino-higgsino mixtures. The second
lightest neutralino $\tilde \chi_{2}^{0}$ is almost a pure weak
eigenstate in each of the three scenarios: a zino in scenario (A), a
photino in scenario (B) and a higgsino in scenario (C) with only small
admixtures from other weak eigenstates. Similarly $\tilde \chi_{3}^{0}$
is almost a higgsino in scenarios (A) and (B) and a photino in (C). The
heaviest neutralino $\tilde \chi_{4}^{0}$ finally is almost a pure
higgsino in scenario (A), a zino in scenario (C) and a zino-higgsino
mixture with a rather small photino component in scenario (B). We shall
see that the question which of the neutralinos or charginos will be
produced with the highest rate sensitively  depends on the nature of the
mass eigenstates. Thus it may happen that the cross section for the
production of the heaviest neutralino $\tilde \chi_{4}^{0}$ is by one
order of magnitude higher than that for the lightest one $\tilde
\chi_{1}^{0}$.

For each of these mixing scenarios total production cross sections have
been calculated for three different relations between slepton mass and
squark mass: $m_{\tilde q}=m_{\tilde l}$ in scenarios (A1), (B1) and
(C1) and $m_{\tilde q}=4\cdot m_{\tilde l}$ in scenarios (A2), (B2) and
(C2). Three further scenarios (A'), (B') and (C') with $m_{\tilde
l}=m_{\tilde g}$ and $m_{\tilde q}=1.4 \cdot m_{\tilde l}$ are
motivated by the renormalization group relation eq. (\ref{eqpol}) coupling
the sfermion masses and the gaugino mass parameter $M$ of the MSSM.
Allowing an error of at most 4\% for the sfermion masses this choice
(neglecting the mass difference between left and right-handed as well as
between up and down-type sfermions) is for values of $M$ between 45 GeV
and 450 GeV and $\tan \beta =2$ compatible with the mass relation eq.
(\ref{eqpol}). The value of $\mu $ in scenarios (A'), (B') and (C') is the
same as in scenarios (A), (B) and (C), respectively, whereas $M$ is
variied between 45 GeV and 450 GeV, which corresponds to values of the
gluino mass between $m_{\tilde g}=1.5\cdot m_Z$ and $m_{\tilde
g}=15\cdot m_Z$. Notice that from there in these scenarios both the mass
and the mixing character of the neutralinos and charginos depend on the
respective value of the sfermion mass.

In order to compute the cross sections for ep-scattering, eq.
(\ref{integ}), we fold the amplitudes squared for the parton subprocesses with
the quark distribution functions $q_k (x,\tilde Q^2)$ of Gl\"uck, Reya
and Vogt \cite{GRV} and integrate over the whole phase space.
In contrast to slepton-squark production the momentum transfer squared to
the nucleon $\tilde Q^2$ depends on the respective production mechanism:
$\tilde Q^2 =Q^2=-(p_{q_{out}}-p_{q_{in}})^2 $ for the graphs 1, 3 and
5, whereas for the graphs 2 and 4 one has $\tilde Q^2 ={Q'}^2=-(p_{\tilde
l}-p_{e})^2$. We have, however, numerically checked that in the
kinematic region investigated here $\tilde Q^2 =\frac{1}{2}\left(sx-(m_{\tilde
\nu } + m_{\tilde \chi_i})^2\right)$ is a satisfactory approximation
leading to an error of at most 10\% in the most unfavorable case of
processes dominated by neutralino or chargino exchange. For processes
dominated by gauge boson exchange the error is smaller.

For avoiding divergences arising from photon exchange and also in order
to separate deep inelastic from elastic scattering and exclusive
inelastic processes we impose a cut for the momentum transfer squared
$Q^2$ to the quark with $Q_{cut}^{2}=10(\mbox{GeV})^2$. It is true that
this choice of $Q_{cut}^{2}$ involves some uncertainty for the values of
the total cross sections. We have, however, numerically checked that
processes dominated by the exchange of massive gauge bosons or
neutralinos/charginos are rather insensitive to the actual value of
$Q_{cut}^{2}$. For processes dominated by photon exchange the dependence
on $Q_{cut}^{2}$ of the cross sections is approximately logarithmic (i.e.
for $Q_{cut}^{2}=10/5/2.5$ GeV we get for the production of $\tilde
\chi_{1}^{0}$ and a selectron in scenario (A.2) and $m_{\tilde e}=300$
GeV a cross section from $\sigma= 1.2/1.3/1.4\cdot 10^{-4}$ pb).

For the squark width entering into ${\cal M}_2$ we have taken into
account all contributions from its two body decays. Three body decays are
suppressed in case of our scenarios \cite{HaKa}. The integration was
performed using the monte carlo program {\em vegan}.
\subsection{The Process $ep\rightarrow \tilde e \tilde \chi_{i}^{0}
X$}
{}From all production channels investigated in this paper only associated
production of a selectron and a photino-like LSP $\tilde \chi_{1}^{0}
\simeq \tilde \gamma$ has been examined in detail in the literature
\cite{ammr}.

In figs. 2--5 we show total cross sections $\sigma ( ep \rightarrow
\tilde e \tilde \chi_{i}^{0} X ),\, i=1,\ldots,4 $ for all four
neutralino states as a function of the selectron mass. Since the cross
sections are rather small, we give as examples the numerical results for
scenarios (A1), (A2) and (C1), (C2) only most obviously revealing
some interesting features. Especially we omit graphs for the scenarios
based on the mass relations eq. (\ref{eqpol}). For scenario (A') the cross
sections are smaller than $10^{-2}$ pb and for scenarios (B') and (C')
smaller than $10^{-3}$ pb.

Apart from the region of small selectron mass in scenario (C1) and
(B1) the
cross sections are the largest for photino-like neutralinos, i.e. the
LSP $\tilde \chi_{1}^{0}$ in scenarios (A1) and (A2), $\tilde
\chi_{2}^{0}$ in scenarios (B1) and (B2) and the heavy
neutralino $\tilde \chi_{3}^{0}$ in scenarios (C1) and (C2). In the
case of a photino-like (or zino-like) neutralino the dominating
contributions are from the Feynman graphs 1 and 3 with strong
electron-selectron-photino couplings. In scenarios (A1), (B1) and
(C1) also graph 2 gives remarkable contributions.
The steep ascent in these scenarios
of the cross sections most strongly marked in scenario (C1) originates
from the contribution of the Feynman graph 2 where for $m_{\tilde
q}\ge m_{\tilde \chi_{i}^{0}}$ the squark approaches its mass shell in
the accessible region of phase space. Then neglecting contributions from
all other graphs as well as interference contributions we have $\sigma
(ep\rightarrow \tilde e \tilde \chi_{i}^{0} X) \simeq \sigma
(ep\rightarrow \tilde e \tilde q X)\cdot BR(\tilde q \rightarrow \tilde
\chi_{i}^{0} q)$ as a reasonable estimation for $m_{\tilde q} >
m_{\tilde \chi_{i}^{0}}$.

It is remarkable that in scenario (C1) for $m_{\tilde e} \ge 150 $ GeV
the cross section for the heavy photino-like neutralino $\tilde
\chi_{3}^{0}$ and for $m_{\tilde e} \ge 300$ GeV even that for the
heaviest neutralino with dominating zino component is by one order of
magnitude higher than that for the lightest higgsino-like state $\tilde
\chi_{1}^{0}$ produced via the reaction mechanism of Feynman graph 5
mainly (pur higgsinos are only produced by graph 5, but only in case of
scenarios with at least one higgsino-chargino mixing state, which couples
at both vertices of the exchanged neutralino).

This demonstrates that the question which of the neutralino cross
sections is the dominating one crucially depends on the mixing
properties of the respective states, whereas their mass is of minor
importance.

For comparison we also give in the figs. the cross sections
for selectron-squark production. For $m_{\tilde q}=m_{\tilde e}$ the
biggest of the neutralino cross sections ($\tilde \chi_{1}^{0}$ for
$m_{\tilde e} \simeq 200$ GeV in scenario (A1) and $\tilde
\chi_{3}^{0}$ for $m_{\tilde e}\simeq 250$ GeV in scenario (C1)) is at
best approximately 30\% of that for selectron-squark production. If,
however, the squarks are considerably heavier than the selectrons
($m_{\tilde q} =4\cdot m_{\tilde e}$) than the selectron-squark cross
section rapidly drops for increasing selectron mass, so that for
$m_{\tilde e}\ge 300 $ GeV associated selectron-neutralino production is
the dominating process.
\subsection{The Process $ep\rightarrow \tilde e \tilde
\chi_{i}^{+ } X$}
For the same reasons as for neutralino production we compare in figs.
6, 7 the total cross sections for $\tilde e \tilde \chi_{i}^{+}$-production
(and $\tilde e \tilde \chi_{i}^{-}$-production) in scenarios (A1),
(A2) only with those for $\tilde e\tilde
q$-production. And we also give no results for the scenarios (A'), (B')
and (C'). In scenarios (A1) and (A2) the cross sections are the
largest for wino-like light charginos $\tilde \chi_{1}^{+}$. Similar as
for production of a photino-like neutralino the dominating contributions
are those from Feynman graph 3 and also from graph 2 if
the squark approaches its mass shell, generating the steep ascent of the
cross sections for $\tilde \chi_{1}^{+}$ in scenario (A1)
with $m_{\tilde q}=m_{\tilde e}$. In contrast to neutralino production the
amplitude ${\cal M}_5$ gives significant contributions to the production of
both gaugino-like and higgsino-like charginos. Being of minor importance for
charginos with dominating wino component, it is the crucial production
mechanism for higgsino-like charginos, i.e. $\tilde \chi_{2}^{+}$ in
scenario (A1), (A2) and $\tilde \chi_{1}^{+}$ in scenarios (C1),
(C2).

In scenarios (C1) and (C2) a consequence of the interplay between mass
and mixing character is the type of the chargino with the largest cross
section changes with increasing selectron mass (and squark mass).
In scenario (C1) it is
for $m_{\tilde q}=m_{\tilde e}<250$ GeV the higgsino-like light chargino,
whereas for $m_{\tilde q}=
m_{\tilde e}>250 $ GeV it is the wino-like heavy chargino,
which is produced with the highest rate. Vice versa in scenario (C2)
with heavy squarks it is for $m_{\tilde e}<200$ GeV the wino-like
$\tilde \chi_{2}^{+}$ whereas for $m_{\tilde e}>200 $ GeV it is the
higgsino-like $\tilde \chi_{1}^{+}$ which yields the highest cross
section. This shows the changing importance of the contributions from
graph 2 compared to the contributions of graphs 1, 3 and 5.

Similar as for the case of neutralino production in scenario (A1) and
(C1) (for $m_{\tilde q}=m_{\tilde e}$) the cross sections for
selectron-chargino production are at best 30\% of that for
selectron-squark production. For $m_{\tilde q}=4\cdot m_{\tilde e}$,
however, the cross section for $\tilde e \tilde q$-production is rapidly
decreasing with increasing selectron mass so that for $m_{\tilde e} \ge
250$ GeV associate selectron-chargino production is the dominating
process with cross sections approximately one order of magnitude higher
than those for selectron-neutralino production.
\subsection{The Process $ep\rightarrow \tilde e \tilde
\chi_{i}^{- } X$}
Production of charginos $\tilde \chi_{i}^{+}$ and anti-charginos $\tilde
\chi_{i}^{-}$ differs in two substantial features. Firstly the
dominating reaction mechanism for production of a wino-like chargino
$\tilde \chi^-$ is via graph 1 instead of graph 3 for a wino-like
chargino $\tilde \chi^+$. Secondly only $d$ valence quarks are
contributing to $\tilde e \tilde \chi^-$-production instead of $u$
valence quarks for $\tilde e\tilde \chi^+$-production. Thus considerably
differences are to be expected for the respective cross sections in
figs. 6, 7. In the case of higgsino-like charginos ${\cal M}_5$ is the
dominating amplitude for both $\tilde \chi^+$ and $\tilde
\chi^-$-production. Again for $m_{\tilde q}=m_{\tilde e}$ one obtains
considerably contributions from squark exchange in amplitude ${\cal
M}_2$ generating in scenario (C1) a rise of the cross section for
$\tilde \chi_{2}^{-}$. For the wino-like $\tilde \chi_{1}^{-}$ in
scenario (A1) this is suppressed by the surmounting contribution of the
amplitude ${\cal M}_1$.

Apart from the region $m_{\tilde e} \ge 250$ GeV in scenario (C1) the
cross section for $\tilde e \tilde \chi_{i}^{-}$-production are larger
than those for $\tilde e \tilde \chi_{i}^{+}$-production and
particularly for large selectron and squark masses higher than those for
$\tilde e \tilde q$-production.
\subsection{The Process $ep\rightarrow \tilde \nu \tilde
\chi_{i}^{0} X$}
For these production channels we compare in figs. 8--14 beside the cross
sections for scenarios (A1), (B1), (C1) and (A2), (B2), (C2) also
that for scenario (A') (based on the mass relation eq. (\ref{eqpol})) with
the cross section for $\tilde \nu \tilde q$-production. Again we omit
figs. for scenarios (B') with cross sections below $10^{-3}$ pb and (C')
with cross sections below $10^{-2}$ pb.

Due to the large $W$-couplings in the Feynman graphs 1, 3 and 5 in all
our scenarios the cross sections are the largest for neutralinos with a
dominating zino component, i.e. $\tilde \chi_{2}^{0}$ in scenarios (A1)
and (A2), $\tilde \chi_{1}^{0}$ and $\tilde \chi_{4}^{0}$
in scenarios (B1) and (B2) and
$\tilde \chi_{4}^{0}$ in scenarios (C1) and (C2). It is noticeable
that in scenarios (B1) and (C1), (C2) it is the heaviest neutralino
which is produced with the highest rate. Especially in scenario (C2)
for $m_{\tilde \nu}> 150$ GeV the cross section for the heaviest
neutralino $\tilde \chi_{4}^{0}$ is one order of magnitude larger than
those for the lighter neutralino states.

Since pure higgsinos would solely be produced via the mechanism of
Feynman graph 5 the cross sections in scenarios (A)
for the heavy higgsino-like states
$\tilde \chi_{3}^{0}$ and $\tilde \chi_{4}^{0}$, respectively, are
nearly equal for $m_{\tilde q}=m_{\tilde \nu}$ and $m_{\tilde
q}=4\cdot m_{\tilde \nu}$. In scenarios (C) the same holds for  the light
higgsino-like state $\tilde \chi_{1}^{0}$, whereas the small zino
component of the higgsino-like neutralino $\tilde \chi_{2}^{0}$ gives
rise to the rather different cross sections for scenarios (C1) and
(C2).

In scenario (A') the mass relation eq. (\ref{eqpol}) couples the gaugino
mass parameter $M$ and the sfermion masses and the neutralino masses as
well as their couplings are varying with increasing sneutrino masses.
This is the reason for the numerous crossings of the cross sections in
fig. 10 for scenario (A').

In all scenarios examined neutralinos with a large zino component are
produced with remarkably large cross sections comparable to or even
considerably larger than those for sneutrino-squark production and also
larger than selectron-squark production. Thus associated
sneutrino-neutralino production appears to provide an attractive channel
in the search for supersymmetric events at LEP$\otimes $LHC. The
importance of this channel is, however, somewhat reduced by the complex
decay patterns of the particles produced leading for a sizeable fraction
of events to final states with many hadronic jets unfavorable for a
clearly visible supersymmetric signal.
\subsection{The Process $ep\rightarrow \tilde \nu \tilde
\chi_{i}^{-} X$}
Due to the dominance of the contributions from
gauge boson exchange in graphs 1, 3 and 5 the numerical results are
nearly identical for $m_{\tilde q}=m_{\tilde \nu}$ and $m_{\tilde
q}=4\cdot m_{\tilde \nu}$. We therefore give
the results for scenarios (A1), (B1) and (C1) and (A'), (B') and
(C') only. In contrast
to $\tilde \nu \tilde q$-production all partons are contributing to
$\tilde \nu \tilde \chi_{i}^{-}$-production.This together with the strong
Z-couplings in graphs 1, 3 and 5 and the photon couplings in graphs 3 and 5
leads for all scenarios to cross sections for sneutrino-chargino
production being between one and two orders of magnitude
larger than those for $\tilde \nu \tilde q(\tilde e\tilde q ) $-production.
The cross section is the highest for the light wino like chargino $\tilde
\chi_{1}^{-}$ in scenario (A1) attaining values between 1pb and 10pb
(for $m_{\tilde \nu}<400 $ GeV) but even for the heavy wino like chargino
$\tilde \chi_{2}^{-}$ in (C1) it is considerably larger than that for
$\tilde \nu \tilde q$-production and also larger than that for the light
higgsino like state $\tilde \chi_{1}^{-}$, with substantial
contributions from ${\cal M}_5$ only. Fig. 17 demonstrates once more the
crucial importance of gaugino higgsino mixing. Both charginos being
mixtures with considerably different masses, the somewhat larger wino
component of the heavier one suffices to raise its cross section to
nearly the same magnitude of that for the lighter one.

In contrast to all other processes discussed in the proceeding sections,
also for scenarios (A'), (B') and (C') the cross sections for $\tilde \nu
\tilde \chi_{i}^{-}$-production are larger than $10^{-2}$ pb in a wide
range of parameter space and considerably larger than those for $\tilde
\nu \tilde q $-production. Notice that in scenarios (A'), (B') and (C') the
mass specified in figs. 16, 18 and 20 as well as the coupling character
of both charginos are varying
with increasing sneutrino mass: For the lower values of $m_{\tilde \nu}$
the light chargino is more wino like, whereas the heavy chargino is
higgsino like. The situation changes with increasing sneutrino mass so
that the light chargino becomes more and more higgsino like whereas the
heavy one becomes more and more wino like. Simultaneously the mass of
the heavy chargino is rapidly increasing whereas that of the light
chargino asymptotically approaches the value $|\mu |$. This interplay
between mass and mixing character produces the two crossings of the
cross sections in figs. 16, 18 and 20.
\section{Signatures}
In order to work out suitable signatures for signals from associate
slepton-neutralino/chargino production it is indispensable to include
the decay of these particles as well as a discussion of the competing
standard model background. Here we shall restrict ourselves to some
remarks, postponing a more detailed discussion of signatures and
background to a subsequent paper.

Light supersymmetric particles decay directly into the lightest neutralino
(which is assumed to be the lightest supersymmetric particle LSP and
stable) and fermions, whereas heavy sparticles decay over complex cascades
ending at the LSP. These cascade decays of heavy sparticles have two
important consequences. On the one hand they will lead to events with
besides one or several leptons, jets and missing energy one or two $W$ or
$Z$ bosons in the final state \cite{tata}.
On the other hand they can significantly
enhance the possible signals of the respective process \cite{cuyp}. The actual
decay patterns and the dominant signatures will, however, sensitively
depend on the supersymmetric parameters and the slepton mass. Thus in
scenario (A1) and (A2) where the cross sections are the biggest for the
processes $ep\rightarrow \tilde \nu \tilde \chi_{2}^{0} X$ and $ep
\rightarrow \tilde \nu \tilde \chi_{1}^{-} X$, and the main decay channels are
$\tilde \nu \rightarrow \nu \tilde \chi_{1}^{-}, \nu \tilde \chi_{2}^{0},
\nu \tilde \chi_{1}^{0}$ and $\tilde \chi_{1}^{-}\rightarrow \tilde
\chi_{1}^{0} e \bar \nu,\, \tilde \chi_{2}^{0}\rightarrow \tilde
\chi_{1}^{0} l \bar l$, the dominant signatures are $
ej\not \! E,\, 2ej\not \! E$ and $3ej \not \! E$ ($j$ denotes an arbitrary
number of jets) for $m_{\tilde l}=100$ GeV. For $m_{\tilde l}=500$ GeV,
where the main production channels are the same but the charginos and
neutralinos decay with a higher probability into lighter charginos/neutralinos
and two quarks, the dominant signatures are $ej\not \! E,\, 2ej\not \! E
$. For scenarios (C1) and (C2) on the other hand
with $ep\rightarrow \tilde \nu \tilde \chi_{4}^{0} X$ and $ep\rightarrow
\tilde \nu \tilde \chi_{2}^{-} X$ as the dominant processes the favored
signatures are $ej \not \! E$ and $Wj\not \! E$ for $m_{\tilde l}=100$ GeV,
arising by the decay of the sneutrino into a lepton and a light
neutralino/chargino
and the decay of the heavy neutralino/chargino either into a
chargino/neutralino and a $W$-boson or into a sfermion-fermion pair.
$eWj\not \! E$ and $e2Wj\not \! E$ are favored signatures
for $m_{\tilde l}=500$ GeV in these scenarios, because the sneutrino also
may decay into a heavy neutralino/chargino and a lepton. For
scenarios (B1) and (B2), finally, with dominant contributions from
$\tilde \nu \tilde \chi_{1}^{-}-$ and $\tilde \nu \tilde \chi_{2}^{-}-
$production the decays of the gaugino-higgsino mixtures $\tilde \chi_{1}^{-
}$ and $\tilde \chi_{2}^{-}$ will lead to final states with both one or
several leptons and one or two gauge bosons as favored signatures.

The most important sources of background are single $W$ and $Z$
production  $ep\rightarrow \nu W X, \nu Z X$ and
$ep\rightarrow eWX,eZX$ followed by the decays $W\rightarrow
l\nu_l,\,Z\rightarrow \nu_l \bar \nu_l$ and $Z\rightarrow l^+ l^-$
giving rise to events with one, two or three charged leptons \cite{baur}.
On the other hand the case of single top production $ep\rightarrow \nu \bar t
bX$ followed by the decay $\bar t\rightarrow \bar b W^- $ gives rise to
the $Wj\not \! E$ configuration and the neutral current process $ep
\rightarrow et\bar t X$ is a source of the background for $e2Wj\not \!
E$ events \cite{ali}. Since, however, the cross section for $t\bar
t$-production is rather small,
one would expect that this is the least dangerous
of the competing standard model backgrounds. Detailed Monte Carlo
studies taking into account the background are needed to asses the
observability of the SUSY signal from associate
slepton-neutralino/chargino production.
\section{Conclusion}
We have analyzed for three representative gaugino-higgsino mixing scenarios
and for different slepton-squark mass ratios
associate slepton-neutralino and
slepton-chargino production at LEP$\otimes $LHC. From all five production
channels
sneutrino-chargino production appears to be the most attractive one.
The cross sections for wino-like light charginos
$\tilde \chi_{1}^{-}$ in scenarios (A) as well as $\tilde \chi_{2}^{-}$ in
scenarios (C) are between one and two orders of
magnitude bigger than those for $\tilde \nu \tilde q$-production: about 0.1 pb
for $m_{\tilde \nu}=500$ GeV and between 1 pb and 10 pb for $m_{\tilde
\nu}=50$ GeV. Similar results are obtained for higgsino-gaugino mixtures
and even for light higgsino-like charginos the cross sections are one order of
magnitude bigger than those for $\tilde \nu \tilde q$-production. Also for
associated production of a zino-like neutralino and a sneutrino the cross
sections are bigger (scenarios (A)) or comparable to (scenarios (C)) those for
$\tilde \nu \tilde q$-production.

For all other production channels, especially for selectron-chargino
production, the situation depends on the squark-slepton
mass ratio and the gaugino-higgsino
mixing scenario. Thus for $m_{\tilde q}=m_{\tilde l}$ the dominating process
is slepton-squark production whereas for  $m_{\tilde q}=4\cdot m_{\tilde l}$
and $m_{\tilde l}>100$ GeV in scenarios (A) ($m_{\tilde l}>200$ GeV in
scenarios (B) and (C)) the cross sections for selectron-squark production are
bigger than those for $\tilde e \tilde q$-production. Similarly for
$m_{\tilde q}=4\cdot m_{\tilde l}$ and $m_{\tilde l}>220$ GeV also
selectron-neutralino production is distinguished by cross sections
larger than those for  $\tilde e \tilde q$-production.

For scenarios (A')--(C') motivated by the mass relation eq. (\ref{eqpol})
only the
sneutrino-chargino cross sections are bigger than $10^{-2}$ pb in a
noticeable range of the parameter space.

The question, which of the neutralinos/charginos will be produced with
the highest rate, depends much more sensitively on the mixing
properties than on their masses. Generally the
production of gaugino-like states is considerably favored so that for
selectron-neutralino production in scenario (C4) it is even the heavy
photino-like neutralino $\tilde \chi_{3}^{0}$ which yields the dominating
cross section. We find the same situation for sneutrino-chargino production:
in scenario (C1) the cross section for the wino-like heavy chargino
$\tilde \chi_{2}^{-}$ is considerably higher than that for the
higgsino-like light chargino $\tilde \chi_{1}^{-}$.

The subsequent decays of the produced sparticles lead to interesting
signatures with up to four charged leptons, hadronic jets and in case of
scenarios (B) and (C) massive gauge bosons. A quantitative analysis of these
signatures and the competing background will be postponed to a subsequent
paper.

The size of the cross section for sneutrino-chargino production,
comparable to or even bigger than that for competing standard model
processes,
let us however suggest, that this process should provide an attractive
channel in the discovery of supersymmetric models at LEP$\otimes $LHC.
\section*{Appendix A}\label{app}
We list in this appendix the complete expressions for the amplitudes
squared for the production channels (a)--(e) convenient for future
computation of differential cross sections, energy spectra etc. The
notation is as in eqs. (\ref{eqeb1})-(\ref{eqeb5}).
For the interference terms we write $({\cal
M}_{i,j})^2 := 2Re({\cal M}_i {\cal M}^{\ast}_{j})$.

Where the expressions are depending on the
polarisations we will write on the lefthand side $a=b$ ($a\neq b$) for
the same (opposite) polarisation of the incoming electron and quark. On
the righthand side $a,b$ occurs in some cases in the formulas with
$a,b=\pm 1 $ for right-, lefthanded electrons and quarks, respectively,
or as an index at the couplings $({\cal O }^{a,b}_{X})_{ij}$. If the
expression is valid for arbitrary polarisations we have suppressed
$a,b$. The $m_{\tilde \chi_i}$ are the eigenvalues of the
neutralino/chargino mixing matrix, respectively, and not the physical
masses and so may be negativ.

The antisymmetric terms -- contractions with $\epsilon$-tensors -- give no
contributions to the total cross section, because they vanish by the phase
space integration.

As for the couplings defined in eqs. (\ref{eqeb1})-(\ref{eqeb5})
we have indicated by a suffix the corresponding Feynman graph:
$(\eta_{e}^{1})_i$ is, for instance, the coupling
$(\eta_{e_a})_i$ in graph 1 and
$(\eta_{q}^{4\ast})_l$  is the complex conjugated from
$(\eta_{q_b}^{\ast})_j$ in graph 4. Depending on the respective process one
has in the amplitudes ${\cal M}_1$, ${\cal M}_3$ and ${\cal M}_5$ to sum
over different contributions from exchange of gauge bosons $X$, given in
table 2. This has for clearness been indicated by a suffix
(${\sum_X}^1$ in the expression for $|{\cal M}_1|^2$, e.g.). The factors
$({\cal P}^{k}_{x})$ etc. refer to the propagator of the particle $x$
exchanged in Feynman graph $k$. To give an example:
\[
 ({\cal P}^{2}_{\tilde \chi})_j =\frac{1}{(\pl-\pe)^2-\mxj^2}.
\]
\begin{eqnarray} \label{exdif1}
\left|{\cal M}_1\right|^2&=&4\left|{\sum_{X}}^1
(f_{e}^{1})_X(f_{q}^{1})_X(\eta_{e}^{1})_{i}
({\cal P}^{1}_{\tilde l})({\cal P}^{1}_{X})\right|^2 \nonumber \\
 & & \Bigm[ 2(\pl \cdot \pqo +\pe \cdot \pqo -\pqo \cdot \pxi )
(\pl \cdot \pqi +\pe \cdot \pqi -\pqi \cdot \pxi )- \nonumber \\
 & &  (m^{2}_{\tilde l}
+2\pe \cdot \pl -2\pl \cdot \pxi +m^{2}_{\tilde \chi_i}-2\pe \cdot \pxi )
\pqi \cdot \pqo \Big] \pe \cdot \pxi \\[0.2cm]
\left|{\cal M}_2\right|^{2}_{a=b}&=&4\left|\sum_j m_{\tilde \chi_j}
(\eta_{e}^{2})_j (\eta_{q}^{2})_j (\eta_{q}^{2})_i
({\cal P}^{2}_{\tilde \chi})_j ({\cal P}^{2}_{\tilde q})\right|^2 \pe
\cdot \pqi \pxi \cdot \pqo \\[0.2cm]
\left|{\cal M}_2\right|^{2}_{a\neq b}&=&4\left|\sum_j (\eta_{e}^{2})_j
(\eta_{q}^{2})_j (\eta_{q}^{2})_i
({\cal P}^{2}_{\tilde \chi})_j ({\cal P}^{2}_{\tilde q})\right|^2 \nonumber \\
 & & \pqo \cdot \pxi
\left(2\pe \cdot \pl (\pl \cdot \pqi -\pe \cdot \pqi )-\pe \cdot \pqi
(m^{2}_{\tilde l}-2\pe \cdot \pl )\right) \\[0.2cm]
\left|{\cal M}_3\right|^{2}_{a=b}&=&16\left| {\sum_{X}}^3
(f_{e}^{3})_X (f_{q}^{3})_X (\eta_{e}^{3})_{i}
({\cal P}_{e}^{3})({\cal P}^{3}_{X})\right|^2 \nonumber \\
 & & \pe \cdot \pqi \Bigm[2(
m^{2}_{\tilde \chi_i}+\pl \cdot \pxi)(\pqo \cdot \pxi + \pl \cdot \pqo )-
  \nonumber \\
 & &  \pqo \cdot \pxi (m^{2}_{\tilde \chi_i}+
2\pl \cdot \pxi +m^{2}_{\tilde l})\Big] \\[0.2cm]
\left|{\cal M}_{3}\right|_{a\neq b}^{2}&=&16\left|{\sum_{X}}^3
(f_{e}^{3})_X (f_{q}^{3})_X (\eta^{3}_{e})_{i}
({\cal P}^{3}_{e})({\cal P}^{3}_{X})\right|^2 \nonumber \\
 & & \pe \cdot \pqo \bigm[2(
m^{2}_{\tilde \chi_i}+\pl \cdot \pxi)(\pqi \cdot \pxi + \pl \cdot \pqi )-
\nonumber \\
 & &  \pqi \cdot \pxi (m^{2}_{\tilde \chi_i}+
2\pl \cdot \pxi +m^{2}_{\tilde l})\big] \\[0.2cm]
\left|{\cal M}_{4}\right|_{a=b}^{2}&=&4\left|\sum_j (\eta_{e}^{4})_j
(\eta_{q}^{4})_j (\eta_{q}^{4})_i
({\cal P}^{4}_{\tilde \chi})_j ({\cal P}^{4}_{\tilde q})\right|^2 \nonumber \\
 & & \pqi \cdot \pxi
\left(2\pe \cdot \pl (\pl \cdot \pqo -\pe \cdot \pqo )-\pe \cdot \pqo (
m^{2}_{\tilde l} -2\pe \cdot \pl )\right) \\[0.2cm]
\left|{\cal M}_4\right|^{2}_{a\neq b}&=&4\left|\sum_j m_{\tilde \chi_j}
(\eta_{e}^{4})_j (\eta_{q}^{4})_j (\eta_{q}^{4})_i
({\cal P}^{4}_{\tilde \chi})_j ({\cal P}^{4}_{\tilde q})\right|^2 \pe
\cdot \pqo \pxi \cdot \pqi \\[0.2cm]
\left|{\cal M}_5\right|^2&=&4\sum_{k,l} {\sum_{X,X'}}^5
\left( (f_{q}^{5})_X (\eta_{e}^{5})_k
({\cal P}^{5}_{X})({\cal P}^{5}_{\tilde \chi})_k
(f_{q}^{5\ast})_{X'}(\eta_{e}^{5\ast})_l
({\cal P}^{5}_{X'})({\cal P}^{5}_{\tilde \chi})_l
\right) \nonumber \\
 & & \Big[-2({\cal O}^{-a}_{X})_{ik}({\cal O}^{-a\ast }_{X'})_{il}
\pe \cdot \pl \big[2 \pqo \cdot \pxi (\pe \cdot \pqi
-\pl \cdot \pqi )+\nonumber \\ & &
2 \pqi \cdot \pxi (\pe \cdot \pqo -\pl \cdot \pqo )
\big] - \nonumber \\
 & & \big[({\cal O}^{-a}_{X})_{ik}({\cal O}^{-a\ast }_{X'})_{il}(m^{2}_{
\tilde l}-
2\pe \cdot \pl )-({\cal O}^{a}_{X})_{ik}({\cal O}^{a\ast }_{X'})_{il}
m_{\tilde \chi_k}m_{\tilde \chi_l}\big]\nonumber \\ & &
(2\pqi \cdot \pxi \pe \cdot \pqo +2\pe \cdot \pqi \pqo \cdot \pxi )+
 \nonumber\\
& & 2m_{\tilde \chi_i}\big[({\cal O}^{a}_{X})_{ik}({\cal O}^{-a\ast }_{X'}
)_{il}m_{\tilde \chi_k}+
({\cal O}^{-a}_{X})_{ik}({\cal O}^{a\ast }_{X'})_{il}m_{\tilde \chi_l}\big]
\pe \cdot \pl \pqi \cdot \pqo - \nonumber \\
 & & 2ab\Big\{-2({\cal O}^{-a}_{X})_{ik}({\cal O}^{-a\ast }_{X'})_{il}
\pe \cdot \pl \big[\pqo \cdot \pxi (\pe \cdot \pqi
-\pl \cdot \pqi ) -\nonumber \\ & & \pqi \cdot \pxi
(\pe \cdot \pqo -\pl \cdot \pqo )\big] + \nonumber \\
 & &  \big[({\cal O}^{-a}_{X})_{ik}({\cal O}^{-a\ast }_{X'})_{il}
(m^{2}_{\tilde l} -2\pe \cdot \pl )+
({\cal O}^{a}_{X})_{ik}({\cal O}^{a\ast }_{X'})_{il}
m_{\tilde \chi_k} m_{\tilde \chi_l}\big]\nonumber \\ & &
(\pqi \cdot \pxi \pe \cdot \pqo -\pe \cdot \pqi \pqo \cdot \pxi )-
m_{\tilde \chi_i}\big[ ({\cal O}^{a}_{X})_{ik}({\cal O}_{X'}^
{-a\ast })_{il}m_{\tilde \chi_k}+ \nonumber \\ & &
({\cal O}^{-a}_{X})_{ik}({\cal O}^{a\ast }_{X'})_{il} m_{\tilde \chi_l}\big]
(\pl \cdot \pqo \pe \cdot \pqi -\pl \cdot \pqi \pe \cdot \pqo )\Big\} \Big]
\end{eqnarray}
\begin{eqnarray}
({\cal M}_{1,2})^{2}_{a=b}&=&-4Re\sum_j {\sum_X}^1 \left((\eta_{e}^{2})_j
(\eta_{q}^{2})_j m_{\tilde \chi_j} (\eta_{q}^{2})_i ({\cal P}^{2}_{\tilde q}
) ({\cal P}^{2}_{\tilde \chi})_j (\eta_{e}^{1 \ast})_i
(f_{e}^{1\ast})_X (f_{q}^{1\ast})_X ({\cal P}^{1}_{X})({\cal P}^{1}_{
\tilde l}) m_{\tilde \chi_i}\right) \nonumber\\
& & \Big[(\pl \cdot \pqo +\pe \cdot \pqo -\pxi \cdot \pqo)\pe \cdot \pqi +
\nonumber \\ & &
(\pl \cdot \pqi +\pe \cdot \pqi -\pxi \cdot \pqi)\pe \cdot \pqo -
\nonumber \\
& & (\pe \cdot \pl -\pe \cdot \pxi)\pqi \cdot \pqo \Big]- \nonumber \\
& & 8Im\sum_j {\sum_X}^1 \left((\eta_{e}^{2})_j
(\eta_{q}^{2})_j m_{\tilde \chi_j} (\eta_{q}^{2})_i ({\cal P}^{2}_{\tilde q}
) ({\cal P}^{2}_{\tilde \chi})_j (\eta_{e}^{1 \ast})_i
(f_{e}^{1\ast})_X (f_{q}^{1\ast})_X ({\cal P}^{1}_{X})({\cal P}^{1}_{
\tilde l}) m_{\tilde \chi_i}\right)
\nonumber \\ & & a\cdot \epsilon^{\mu \nu \sigma \tau }p_{e}^{\mu}
p_{\tilde l}^{\nu}p_{q_{in}}^{\sigma}p_{q_{out}}^{\tau} \\[0.2cm]
({\cal M}_{1,2})^{2}_{a\neq b}&=&-8Re\sum_j {\sum_X}^1 \left((\eta_{e}^{2})_j
(\eta_{q}^{2})_j (\eta_{q}^{2})_i ({\cal P}^{2}_{\tilde q}
) ({\cal P}^{2}_{\tilde \chi})_j (\eta_{e}^{1 \ast})_i
(f_{e}^{1\ast})_X (f_{q}^{1\ast})_X ({\cal P}^{1}_{X})({\cal P}^{1}_{
\tilde l}) \right) \nonumber\\
& & \Big[ (m_{\tilde \chi_i}^{2}-2\pe \cdot \pqo )(\pqi \cdot \pqo \pe \cdot
\pxi + \nonumber \\ & &
\pqo \cdot \pxi \pe \cdot \pqi -\pe \cdot \pqo \pqi \cdot \pxi )-
 \nonumber\\
& & 2\pqo \cdot \pxi ( \pqi \cdot \pxi \pe \cdot \pqo +\pe \cdot \pxi \pqi
\cdot \pqo -\pqo \cdot \pxi \pe \cdot \pqi )+\nonumber \\
& & 2m^{2}_{\tilde \chi_i}\pe \cdot
\pqo ( \pqi \cdot \pqo -\pe \cdot \pqi + \pxi \cdot \pqi )- \nonumber \\
 & & 4\pqi \cdot \pxi \pe \cdot \pxi \pqo \cdot \pxi +\nonumber \\
& & 2\pe \cdot \pxi
(\pe \cdot \pqo \pxi \cdot \pqi + \pqo \cdot \pxi \pe \cdot \pqi - \pqi
\cdot \pqo \pe \cdot \pxi ) \Big] +\nonumber \\
& & 8Im\sum_j {\sum_X}^1 \left((\eta_{e}^{2})_j
(\eta_{q}^{2})_j (\eta_{q}^{2})_i ({\cal P}^{2}_{\tilde q}
) ({\cal P}^{2}_{\tilde \chi})_j (\eta_{e}^{1 \ast})_i
(f_{e}^{1\ast})_X (f_{q}^{1\ast})_X ({\cal P}^{1}_{X})({\cal P}^{1}_{
\tilde l}) \right) \nonumber\\
& & a(2 \pl \cdot \pe -\ml^2)\cdot \epsilon^{\mu \nu \sigma \tau}
\pe^{\mu} \pl^{\nu} \pqi^{\sigma} \pqo^{\tau} \\[0.2cm]
({\cal M}_{1,3})^{2}&=&8Re {\sum_X}^1 {\sum_{X'}}^3
\left( (f_{e}^{1})_X (f_{q}^{1})_X (\eta_{e}^{1})_i
({\cal P}_{X}^{1})({\cal P}_{\tilde l}^{1})(f_{e}^{3\ast})_{X'}
(f_{q}^{3\ast})_{X'}(\eta_{e}^{3\ast})_i ({\cal P}_{X'}^{3})
({\cal P}_{e}^{3})\right) \nonumber \\
& & \Big[ \bigm[ \pe \cdot \pqi (m_{\tilde \chi_i}^{2} +\pl \cdot \pxi )+
\pe \cdot \pxi (\pqi \cdot \pxi +\pl \cdot \pqi ) -
\nonumber \\ & & (\pe \cdot \pxi +\pe \cdot
\pl )\pqi \cdot \pxi \big]
(\pl \cdot \pqo +\pe \cdot \pqo  -\pqo \cdot \pxi) + \nonumber \\ & &
(\pl \cdot \pqi + \pe \cdot \pqi - \pqi \cdot \pxi ) \bigm[ \pe \cdot
\pqo (\mxi^2 +\pl \cdot \pxi ) + \nonumber \\ & &
\pe \cdot \pxi (\pqo \cdot \pxi +\pl \cdot
\pqo )-(\pe \cdot \pxi + \pe \cdot \pl )\pqo \cdot \pxi \big]-\nonumber\\
& & \pqi \cdot \pqo \big[ (\pe \cdot \pl -\pe \cdot \pxi )(\mxi^2 +\pl \cdot
\pxi )+ \nonumber \\ & &
\pe \cdot \pxi (\pe \cdot \pxi -\mxi^2 +\ml^2 +\pe \cdot \pl )-
\nonumber \\
 & & (\pe \cdot \pxi +\pe \cdot \pl )(\pl \cdot \pxi +
\pe \cdot \pxi -\mxi^2 )\big]+\nonumber \\
& &  ab \Big\{ \pe \cdot \pxi \pqi \cdot \pxi (\pe \cdot \pxi -\mxi^2 +
\ml^2 +\pe \cdot \pl )- \nonumber \\ & &
\pe \cdot \pqo (\pxi \cdot \pqi + \pl \cdot \pqi )
(\pl \cdot \pxi + \pe \cdot \pxi -\mxi^2 )+ \nonumber \\ & &
(\pe \cdot \pl - \pe \cdot \pxi )(\pqi \cdot \pxi +\pqi \cdot \pl )\pqo
\cdot \pxi -\nonumber \\
& & (\pe \cdot \pl -\pe \cdot \pxi )(\pqo \cdot \pxi +\pqo \cdot
\pl )\pqi \cdot \pxi + \nonumber \\ & &
\pe \cdot \pqi (\pqo \cdot \pxi +\pl \cdot \pqo )(\pl \cdot
\pxi +\pe \cdot \pxi -\mxi^2 )- \nonumber \\
& & \pe \cdot \pqi \pxi \cdot \pqo (\pe \cdot
\pxi -\mxi^2 +\ml^2 + \pe \cdot \pl ) \Big\} \Big] \\[0.2cm]
({\cal M}_{1,4})^{2}_{a=b}&=&-8Re\sum_j {\sum_X}^1 \left((\eta_{e}^{4})_j
(\eta_{q}^{4})_j  (\eta_{q}^{4})_i ({\cal P}^{4}_{\tilde q}
) ({\cal P}^{4}_{\tilde \chi})_j (\eta_{e}^{1 \ast})_i
(f_{e}^{1\ast})_X (f_{q}^{1\ast})_X ({\cal P}^{1}_{X})({\cal P}^{1}_{
\tilde l}) \right) \nonumber\\
& & \Big\{ 4(\pe \cdot \pxi -\pe \cdot \pqi )(\pqi \cdot \pxi -\pe \cdot
\pqi )\pqo \cdot \pxi - \nonumber \\
& & 2\bigm[ \pqo \cdot \pxi
(\pqi \cdot \pxi -\pe \cdot \pqi )+
\pqi \cdot \pqo ( \mxi^2 -\pe \cdot \pxi )-\nonumber\\
& & (\pqo \cdot \pxi -\pe
\cdot \pqo )\pqi \cdot \pxi \big](\pe \cdot \pxi -\pe \cdot \pqi )-\nonumber\\
& & \mxi^2\bigm[ \pe \cdot \pqo (\pqi \cdot
\pxi -\pe \cdot \pqi )+\pe \cdot \pqi (\pqo \cdot \pxi -\pe \cdot \pqo )-
\nonumber \\
& &  \pe \cdot \pxi \pqi \cdot \pqo \big] +
2\pqi \cdot \pxi \bigm[\pe \cdot \pqi (\pqo \cdot \pxi -\pe \cdot
\pqo )+\nonumber \\
& & \pe \cdot \pqo (\pqi \cdot \pxi -\pe \cdot \pqi )-\pqi \cdot \pqo
\pe \cdot \pxi \big] \Big\} \\[0.2cm]
({\cal M}_{1,4})^{2}_{a\neq b}&=&4Re\sum_j {\sum_X }^1\left(
m_{\tilde \chi_j}(\eta_{e}^{4})_j
(\eta_{q}^{4})_j  (\eta_{q}^{4})_i ({\cal P}^{4}_{\tilde q}
) ({\cal P}^{4}_{\tilde \chi})_j (\eta_{e}^{1 \ast})_i
(f_{e}^{1\ast})_X (f_{q}^{1\ast})_X ({\cal P}^{1}_{X})({\cal P}^{1}_{
\tilde l}) \right) \nonumber\\
& & m_{\tilde \chi_i}
\Big[ (\pl \cdot \pqo +\pe \cdot \pqo -\pxi \cdot \pqo )\pe \cdot \pqi+
\nonumber \\ & &
(\pl \cdot \pqi +\pe \cdot \pqi -\pxi \cdot \pqi )\pe \cdot \pqo -
\nonumber \\
& & \pqi \cdot \pqo (\pe \cdot \pl -\pe \cdot \pxi ) \Big] \\
[0.2cm]
({\cal M}_{1,5})^{2}&=&16Re\sum_j {\sum_X}^5 {\sum_{X'}}^1 \Big(
(\eta_{e}^{5})_j (f_{q}^{5})_X ({\cal P}^{5}_{\tilde \chi})_j
({\cal P}^{5}_{X})(\eta_{e}^{1 \ast})_i
(f_{e}^{1\ast})_{X'}(f_{q}^{1\ast})_{X'}({\cal P}^{1}_{X'})({\cal P}^{1}_{
\tilde l}) \nonumber\\
& &  \Big[({\cal O}^{a}_{X})_{ij}
\mxi \mxj (\pl \cdot \pqo \pe \cdot \pqi +\pl \cdot \pqi
\pe \cdot \pqo -\nonumber \\ & &
\pqi \cdot \pqo \pl \cdot \pe ) -
({\cal O}^{-a}_{X})_{ij} \bigm[ \pe \cdot \pxi (2\pl
\cdot \pqi \pl \cdot \pqo - \pqi \cdot \pqo \ml^2 )+ \nonumber \\
& & \pe \cdot \pl (\pl \cdot \pqo \pqi \cdot \pxi +\pl \cdot \pqi
\pxi \cdot \pqo -
\pqi \cdot \pqo \pl \cdot \pxi )-\nonumber \\
& & \pl \cdot \pxi (\pqo \cdot
\pl \pqi \cdot \pe +\pqi \cdot \pl \pqo \cdot \pe -\pqi \cdot \pqo \pe \cdot
\pl ) \big] + \nonumber \\
& & ab({\cal O}^{-a}_{X})_{ij}
\big\{ \pqo \cdot \pxi \pe \cdot \pl \pl \cdot \pqi -\pqo \cdot \pxi
\pe \cdot \pqi \ml^2 -\nonumber \\ & &
\pl \cdot \pxi \pe \cdot \pqo \pl \cdot \pqi+
\pl \cdot
\pxi \pe \cdot \pqi \pl \cdot \pqo +\nonumber \\ & &
\pxi \cdot \pqi \pe \cdot \pqo \ml^2 -
\pxi \cdot \pqi \pe \cdot \pl \pl \cdot \pqo \big\} \Big] \Big) \\[0.2cm]
({\cal M}_{2,3})^{2}_{a=b}&=&-16Re\sum_j {\sum_X}^3
\left(m_{\tilde \chi_j}(\eta_{e}^{2})_j
(\eta_{q}^{2})_j  (\eta_{q}^{2})_i ({\cal P}^{2}_{\tilde q}
) ({\cal P}^{2}_{\tilde \chi})_j (\eta_{e}^{3 \ast})_i
(f_{e}^{3\ast})_X (f_{q}^{3\ast})_X ({\cal P}^{3}_{X})({\cal P}^{3}_{e})
\right) \nonumber\\ & & m_{\tilde \chi_i}
\pe \cdot \pqi (\pl \cdot \pqo + \pxi \cdot \pqo ) \\[0.2cm]
({\cal M}_{2,3})^{2}_{a\neq b}&=&-16Re\sum_j {\sum_X}^3
\left((\eta_{e}^{2})_j
(\eta_{q}^{2})_j (\eta_{q}^{2})_i ({\cal P}^{2}_{\tilde q}
) ({\cal P}^{2}_{\tilde \chi})_j (\eta_{e}^{3 \ast})_i
(f_{e}^{3\ast})_X (f_{q}^{3\ast})_X ({\cal P}^{3}_{X})({\cal P}^{3}_{e})
\right) \nonumber\\
& & \Big[ ( -\pqi \cdot \pqo -\pqi \cdot \pxi )\bigm[ \pqo \cdot \pxi
(\pe \cdot \pqi -\pe \cdot \pqo ) + \nonumber \\ & &
\pe \cdot \pqo (\mxi^2 +\pl \cdot \pxi )-
(\pe \cdot \pqo + \pqi \cdot \pqo ) \pe \cdot \pxi \big]+ \nonumber \\
& & (\pqo \cdot \pxi \pe \cdot \pqi +\pe \cdot \pqo \pqi \cdot \pxi
-\pqi \cdot \pqo \pe \cdot \pxi )\nonumber\\
& & (\pe \cdot \pqo +\pqi \cdot \pqo + \pqi \cdot
\pxi +\pe \cdot \pxi -\pqo \cdot \pxi )- \nonumber \\
& & 2\pqo \cdot \pxi \pe \cdot \pqi
\pe \cdot \pqo -\pe \cdot \pqi \pe \cdot \pqo \mxi^2 \Big]- \nonumber \\
& & 16Im\sum_j {\sum_X}^3 \left((\eta_{e}^{2})_j
(\eta_{q}^{2})_j (\eta_{q}^{2})_i ({\cal P}^{2}_{\tilde q}
) ({\cal P}^{2}_{\tilde \chi})_j (\eta_{e}^{3 \ast})_i
(f_{e}^{3\ast})_X (f_{q}^{3\ast})_X ({\cal P}^{3}_{X})({\cal P}^{3}_{e})
\right) \nonumber\\
& & a(\pe \cdot \pl - \pl \cdot \pqo)\cdot \epsilon^{\mu \nu \sigma \tau }
\pe^{\mu} \pl^{\nu} \pqi^{\sigma} \pqo^{\tau} \\[0.2cm]
({\cal M}_{2,4})_{a=b}^{2}&=&4 Re\left(\sum_{k} m_{\tilde \chi_{k}}
(\eta_{e}^{2})_{k} (\eta_{q}^{2})_{k} (\eta_{q}^{2})_i
({\cal P}^{2}_{\tilde \chi})_{k} ({\cal P}^{2}_{\tilde q})\right)\left(
\sum_{l} (\eta_{e}^{4\ast })_{l} (\eta_{q}^{4\ast })_{l} (\eta_{q}^{4\ast })_i
({\cal P}^{4}_{\tilde \chi})_l ({\cal P}^{4}_{\tilde q})\right)\nonumber \\
& & \mxi \bigm[ \pe \cdot \pqi (\pe \cdot \pqo -\pl \cdot \pqo )-
\nonumber \\ & & \pe \cdot \pqo (\pe \cdot \pqi -\pl \cdot \pqi )-
\pqi \cdot \pqo \pe \cdot \pl \big]+ \nonumber \\
& & 4 Im\left(\sum_{k} m_{\tilde \chi_{k}}
(\eta_{e}^{2})_{k} (\eta_{q}^{2})_{k} (\eta_{q}^{2})_i
({\cal P}^{2}_{\tilde \chi})_{k} ({\cal P}^{2}_{\tilde q})\right)\left(
\sum_{l} (\eta_{e}^{4\ast })_{l} (\eta_{q}^{4\ast })_{l} (\eta_{q}^{4\ast })_i
({\cal P}^{4}_{\tilde \chi})_l ({\cal P}^{4}_{\tilde q})\right)\nonumber \\
& & a\cdot \epsilon^{\mu \nu \sigma \tau } \pe^{\mu} \pl^{\nu}
\pqi^{\sigma} \pqo^{\tau} \\[0.2cm]
({\cal M}_{2,4})_{a\neq b}^{2}&=&4 Re\left(\sum_{k}
(\eta_{e}^{2})_{k} (\eta_{q}^{2})_{k} (\eta_{q}^{2})_i
({\cal P}^{2}_{\tilde \chi})_{k} ({\cal P}^{2}_{\tilde q})\right)\left(
\sum_{l} m_{\tilde \chi_{l}}(\eta_{e}^{4\ast })_{l}
(\eta_{q}^{4\ast })_{l} (\eta_{q}^{4\ast })_i
({\cal P}^{4}_{\tilde \chi})_l ({\cal P}^{4}_{\tilde q})\right)\nonumber \\
& & \mxi \bigm[ \pe \cdot \pqo (\pe \cdot \pqi -\pl \cdot \pqi )-
\nonumber \\ & &   \pe \cdot \pqi
(\pe \cdot \pqo -\pl \cdot \pqo )-\pqi \cdot \pqo \pe \cdot \pl \big]
+ \nonumber \\ & &
4 Im\left(\sum_{k}
(\eta_{e}^{2})_{k} (\eta_{q}^{2})_{k} (\eta_{q}^{2})_i
({\cal P}^{2}_{\tilde \chi})_{k} ({\cal P}^{2}_{\tilde q})\right)\left(
\sum_{l} m_{\tilde \chi_{l}}
(\eta_{e}^{4\ast })_{l} (\eta_{q}^{4\ast })_{l} (\eta_{q}^{4\ast })_i
({\cal P}^{4}_{\tilde \chi})_l ({\cal P}^{4}_{\tilde q})\right)\nonumber \\
& & a\cdot \epsilon^{\mu \nu \sigma \tau } \pe^{\mu} \pl^{\nu}
\pqi^{\sigma} \pqo^{\tau} \\[0.2cm]
({\cal M}_{2,5})_{a=b}^{2}&=&-8Re\sum_{k,l} {\sum_X}^5
(\eta_{e}^{2})_{k} (\eta_{q}^{2})_{k}
(\eta_{q}^{2})_i ({\cal P}^{2}_{\tilde \chi})_{k} ({\cal P}^{2}_{\tilde q})
(f_{q}^{5\ast })_X (\eta_{e}^{5\ast })_l ({\cal P}^{5}_{
X})({\cal P}^{5}_{\tilde \chi})_l \nonumber \\
& & \Bigm[ 2 ({\cal O}^{a\ast}_{X})_{il}
m_{\tilde \chi_k}m_{\tilde \chi_l}\pe \cdot \pqi \pxi \cdot
\pqo -({\cal O}^{-a\ast}_{X})_{il}
\mxi m_{\tilde \chi_k}(\pqo \cdot \pe \pqi \cdot \pl -
\nonumber \\ & & \pl \cdot
\pqo \pe \cdot \pqi -\pqi \cdot \pqo \pl \cdot \pe )\Big] \\[0.2cm]
({\cal M}_{2,5})_{a\neq b}^{2}&=&-8aRe\sum_{k,l} {\sum_X}^5
(\eta_{e}^{2})_{k} (\eta_{q}^{2})_{k}
(\eta_{q}^{2})_i ({\cal P}^{2}_{\tilde \chi})_{k} ({\cal P}^{2}_{\tilde q})
(f_{q}^{5\ast })_X (\eta_{e}^{5\ast })_l ({\cal P}^{5}_{
X})({\cal P}^{5}_{\tilde \chi})_l \nonumber \\
& & \Bigm[({\cal O}^{a\ast}_{X})_{il}
\mxi m_{\tilde \chi_l}(\pqo \cdot \pqi \pe \cdot \pl -\pl \cdot
\pqo \pe \cdot \pqi +\pe \cdot \pqo \pl \cdot \pqi )- \nonumber \\ & &
2 ({\cal O}^{-a\ast}_{X})_{il}
\pqo \cdot \pxi (\ml^2 \pe \cdot \pqi -2\pl \cdot \pqi \pl \cdot
\pe )\Big]- \nonumber \\
& & 8Im\sum_{k,l} {\sum_X}^5
(\eta_{e}^{2})_{k} (\eta_{q}^{2})_{k}
(\eta_{q}^{2})_i ({\cal P}^{2}_{\tilde \chi})_{k} ({\cal P}^{2}_{\tilde q})
(f_{q}^{5\ast })_X (\eta_{e}^{5\ast })_l ({\cal P}^{5}_{
X})({\cal P}^{5}_{\tilde \chi})_l \nonumber \\
& & 2a ({\cal O}^{a\ast}_{X})_{il}
\mxi m_{\tilde \chi_l} \cdot \epsilon^{\mu \nu \sigma \tau} \pe^{\mu} \pl^{
\nu} \pqi^{\sigma} \pqo^{\tau} \\[0.2cm]
({\cal M}_{3,4})^{2}_{a=b}&=&16Re\sum_j {\sum_X}^3
\left((\eta_{e}^{4})_j
(\eta_{q}^{4})_j  (\eta_{q}^{4})_i ({\cal P}^{4}_{\tilde q}
) ({\cal P}^{4}_{\tilde \chi})_j (\eta_{e}^{3 \ast})_i(f_{e}^{3\ast})_X
(f_{q}^{3\ast})_X ({\cal P}^{3}_{X})({\cal P}^{3}_{e}) \right)
\nonumber\\
& & \Bigm[\pe \cdot \pqi \pqo \cdot \pxi (2\pe \cdot \pxi - \pe \cdot \pqi )
-\pe \cdot \pqi \pe \cdot \pxi \pqi \cdot \pqo -
\nonumber \\
& & \bigm[\pqi \cdot \pqo (\pe \cdot
\pxi -2\pe \cdot \pqi )+\pe \cdot \pqi \pqo \cdot \pxi -\pe \cdot \pqo \pqi
\cdot \pxi \big] \nonumber \\ & & (\pe \cdot \pxi \pqi \cdot \pxi )
-\pe \cdot \pqi \pe \cdot \pqo (\mxi^2 -\pqi \cdot \pxi )
\Big] \\[0.2cm]
({\cal M}_{3,4})^{2}_{a\neq b}&=&16Re\sum_j {\sum_X}^3 \left((\eta_{e}^{4})_j
(\eta_{q}^{4})_j m_{\tilde \chi_j} (\eta_{q}^{4})_i ({\cal P}^{4}_{\tilde q}
) ({\cal P}^{4}_{\tilde \chi})_j (\eta_{e}^{3 \ast})_i
(f_{e}^{3\ast})_X (f_{q}^{3\ast})_X ({\cal P}^{3}_{X})({\cal P}^{3}_{e})
\right) \nonumber\\
& & m_{\tilde \chi_i}
\left( \pe \cdot \pqo (\pl \cdot \pqi +\pxi \cdot \pqi )\right) \\[0.2cm]
({\cal M}_{3,5})^{2}&=&8Re\sum_j {\sum_X}^5 {\sum_{X'}}^3
\bigg((\eta_{e}^{5})_j (f_{q}^{5})_X ({\cal P}^{5}_{\tilde \chi})_j
({\cal P}^{5}_{X})(\eta_{e}^{3 \ast})_i
(f_{e}^{3\ast})_{X'}(f_{q}^{3\ast})_{X'}({\cal P}^{3}_{X'})({\cal P}^{3}_{e})
 \nonumber\\
& & \Big[ ({\cal O}^{-a}_{X})_{ij}\bigm[ 4\pe
\cdot \pl (\pxi \cdot \pqo \pl \cdot \pqi +\pxi \cdot \pqi \pl \cdot \pqo )
+  \nonumber \\
& &  2\mxi^2\pe \cdot \pl \pqo \cdot \pqi -2\ml^2 (\pxi \cdot \pqo
\pe \cdot \pqi +\pxi \cdot \pqi \pe \cdot \pqo )\big]+ \nonumber \\
& & 2\mxi \mxj ({\cal O}^{a}_{X})_{ij}
\bigm[ \pe \cdot \pqi (\pxi \cdot \pqo +\pl \cdot \pqo )+\nonumber
\\ & & \pe \cdot \pqo (
\pxi \cdot \pqi +\pl \cdot \pqi )\big] -\nonumber \\
& & ab\Big\{ ({\cal O}^{-a}_{X})_{ij} \big[ 4\pe \cdot
\pl (\pxi \cdot \pqo \pl \cdot \pqi -\pxi \cdot \pqi \pl \cdot \pqo )+
\nonumber \\ & &
2\mxi^2 (\pqo \cdot \pe \pqi \cdot \pl - \pe \cdot \pqi \pl \cdot \pqo )
-2\ml^2 (\pxi \cdot \pqo \pe \cdot \pqi -\nonumber\\ & &
\pxi \cdot \pqi \pe \cdot \pqo ) \big]+
2({\cal O}^{a}_{X})_{ij}\mxi \mxj \big[ (
\pxi \cdot \pqo +\pl \cdot \pqo )\pe \cdot \pqi -\nonumber \\
& & (\pxi \cdot \pqi +
\pl \cdot \pqi )\pe \cdot \pqo \big] \Big\}-
\nonumber \\ & & 2\Big( ({\cal O}^{-a}_{X})_{ij} \big[
\pqo \cdot \pl (\pxi \cdot \pe \pqi \cdot \pl +\pe \cdot \pl \pqi \cdot
\pxi -\pl \cdot \pxi \pe \cdot \pqi )+  \nonumber \\
& &  \pl \cdot \pqi (\pe \cdot \pxi \pl \cdot \pqo +\pe \cdot \pl
\pqo \cdot \pxi -\pxi \cdot \pl \pe \cdot \pqo )-\nonumber \\
& & \ml^2 \pqi \cdot \pqo \pe  \cdot \pxi \big] -
ab({\cal O}^{-a}_{X})_{ij}\big\{
\pqo \cdot \pxi \pl \cdot \pe \pqi \cdot \pl -\nonumber \\ & &
\ml^2 \pqo \cdot \pxi \pe \cdot
\pqi -\pe \cdot \pqo \pl \cdot \pxi \pl \cdot \pqi+
\ml^2 \pe \cdot \pqo \pqi \cdot \pxi +\nonumber \\ & &
\pl \cdot
\pqo \pl \cdot \pxi \pe \cdot \pqi -\pl \cdot \pqo \pl \cdot \pe \pqi \cdot
\pxi  \big\} \Big) \Big] \bigg) \\[0.2cm]
({\cal M}_{4,5})_{a=b}^{2}&=&8Re\sum_{k,l} {\sum_X}^5
(\eta_{e}^{4})_{k} (\eta_{q}^{4})_{k}
(\eta_{q}^{4})_i ({\cal P}^{4}_{\tilde \chi})_{k} ({\cal P}^{4}_{\tilde q})
(f_{q}^{5\ast })_X(\eta_{e}^{5\ast })_l ({\cal P}^{5}_{
X})({\cal P}^{5}_{\tilde \chi})_l \nonumber \\
& & \Big[ ({\cal O}^{a\ast}_{X})_{il}\mxi m_{\tilde \chi_l}(\pe \cdot
\pl \pqi \cdot \pqo +\pe \cdot \pqi \pl \cdot \pqo -\pe \cdot \pqo \pl
\cdot \pqi )+\nonumber \\ & &
({\cal O}^{-a\ast}_{X})_{il}( 4\pe \cdot \pl \pqi \cdot \pxi \pl
\cdot \pqo -2\ml^2 \pe \cdot \pqo \pqi \cdot \pxi )   \Big] \\[0.2cm]
\label{exdif2}
({\cal M}_{4,5})_{a\neq b}^{2}&=&8Re\sum_{k,l} {\sum_X}^5
(\eta_{e}^{4})_{k} (\eta_{q}^{4})_{k}
(\eta_{q}^{4})_i ({\cal P}^{4}_{\tilde \chi})_{k} ({\cal P}^{4}_{\tilde q})
(f_{q}^{5\ast })_X (\eta_{e}^{5\ast })_l ({\cal P}^{5}_{
X})({\cal P}^{5}_{\tilde \chi})_l \nonumber \\
& & \Big[ ({\cal O}^{-a\ast}_{X})_{il}\mxi m_{\tilde \chi_k} (\pe \cdot
\pl \pqi \cdot \pqo -\pe \cdot \pqi \pl \cdot \pqo +\pe \cdot \pqo
\pl \cdot \pqi )+ \nonumber\\ & &  2
({\cal O}^{a\ast}_{X})_{il}m_{\tilde \chi_k}m_{\tilde \chi_l}
\pqi \cdot \pxi \pe \cdot \pqo \Big]
\end{eqnarray}
\section*{Appendix B}
In this appendix we give the parametrication of the momenta as well as the
limits for the phase space integration.
We parametrize the momenta of the particles in the
electron quark center-of-mass system as follows:
\begin{displaymath}
\pe=E_{cm}^{eq}\left(\begin{array}{c}
1\\ \sin\vartheta \cos\varphi \\ \sin\vartheta \sin\varphi \\
\cos\vartheta
\end{array} \right) \qquad \pqi=E_{cm}^{eq}\left(\begin{array}{c}
1\\ -\sin\vartheta \cos\varphi \\ -\sin\vartheta \sin\varphi \\
-\cos\vartheta
\end{array} \right)
\end{displaymath}
\begin{displaymath}
\pqo=E_{q_{out}}\left(\begin{array}{c}
1\\ 0 \\ 0 \\ 1 \end{array} \right) \qquad
\pl=\left(\begin{array}{c}
E_{\tilde l}\\ \sqrt{E_{\tilde l}^{2}-\ml^2} \sin\theta \cos\phi \\
\sqrt{E_{\tilde l}^{2}-\ml^2}\sin\theta \sin\phi \\
\sqrt{E_{\tilde l}^{2}-\ml^2}\cos\theta
\end{array} \right)
\end{displaymath}
\begin{displaymath}
\pxi=\left(\begin{array}{c}
2E_{cm}^{eq}-E_{q_{out}}-E_{\tilde l}\\
-\sqrt{E_{\tilde l}^{2}-\ml^2} \sin\theta \cos\phi \\
-\sqrt{E_{\tilde l}^{2}-\ml^2}\sin\theta \sin\phi \\
-\sqrt{E_{\tilde l}^{2}-\ml^2}\cos\theta -E_{q_{out}}
\end{array} \right)
\end{displaymath}
with the momentum of the outgoing quark in direction of the $x_3$-axis,
$\vartheta $ the angle between the electron beam and the $x_3$-axis, $\varphi $
the angle between the $x_1$-$x_3$-plane and the plane fixed by the
$x_3$-axis and the electron beam, $\theta $ the angle between the slepton
momentum and the $x_3$-axis and $\phi $ the angle between the
$x_1$-$x_3$-plane and the plane defined by the momenta of the outgoing
particles. From momentum conservation and the mass shell condition for
$p_{\tilde \chi_i}$ we get
\begin{equation} \label{cosinus}
\cos \theta =\frac{(2E_{cm}^{eq}-E_{q_{out}}-E_{\tilde l})^2-
E_{q_{out}}^{2}-E_{\tilde l}^{2}+m_{\tilde l}^{2}-m_{\tilde \chi_i}^{2
}}{2E_{q_{out}} \sqrt{E_{\tilde l}^{2}-\ml^2}},
\end{equation}
and transform eq. (\ref{integ}) into
\begin{eqnarray} \label{result}
\sigma^{tot}\left(eP\rightarrow \tilde l \tilde \chi_i X \right)&=&
\sum_k \int {\mbox d}x
\frac{1}{256(E_{cm}^{eq})^2 (2\pi )^{5}}\sum_{a,b}
\int  {\mbox d} E_{q_{out}} \int {\mbox d} E_{\tilde l}
\int {\mbox d} (\cos \vartheta ) \int {\mbox d} \varphi \int{\mbox d}
\phi
\times \nonumber \\ & &
q_k(x,Q^2) \left| \sum_n ({\cal M}_n)_{ab} \right|^2 ,
\end{eqnarray}
with the integration limits
\begin{eqnarray*}
0 \leq &\varphi & \leq 2\pi, \\
0 \leq &\phi & \leq 2\pi.
\end{eqnarray*}
The integration limits for $E_{\tilde l},\, E_{q_{out}},\,\vartheta $ and
$x$ are fixed as follows.
{}From $|\cos \theta | \leq 1 $ we obtain
\begin{eqnarray} \label{ElGr}
E_{\tilde l}^{max,min}&=&\frac{2E^{eq}_{cm}-E_{q_{out}}}{2}-\frac
{(2E^{eq}_{cm}-E_{q_{out}})(\mxi^2-\ml^2)}{8E^{eq}_{cm}(E^{eq}_{cm}-
E_{q_{out}})} \pm  \frac{E_{q_{out}}}{2|E^{eq}_{cm}-E_{q_{out}}|}
\times  \nonumber \\
& & \Big[ E_{q_{out}}^{2}+\frac{E_{q_{out}}}{E_{cm}^{eq}}\Big(
\frac{1}{2}(\mxi^2-\ml^2 -4(E_{cm}^{eq})^2)+\ml^2\Big)- \nonumber \\
& &  \ml^2+\frac{1}{(4E_{cm}^{eq})^2}
(\mxi^2-\ml^2-4(E_{cm}^{eq})^2)^2 \Big]^{1/2}.
\end{eqnarray}
{}From eq. (\ref{ElGr}) we get for the upper limit of $E_{q_{out}}$
\[ E_{q_{out}}^{max}=E^{eq}_{cm}\left(1-\frac{(\ml +\mxi )^2}
{(2E^{eq}_{cm})^2}\right) . \]
In order to avoid divergences in the terms containing photon exchange (and
in order to avoid the main part of elastic scattering) we introduce
a cut $Q_{cut}^{2}$ for
\[ Q^2=-(\pl +\pxi -\pe )^2= -(\pqi -\pqo )^2=2E_{cm}^{eq}E_{q_{out}}
(1+\cos \vartheta ) .\]
Then from the condition $Q^2 \geq Q_{cut}^{2}$ the limits for the
$\vartheta$ integration are specified by
\[ \frac{Q^{2}_{cut}}{2E_{cm}^{eP}E_{q_{out}}\sqrt{x}}-1 \leq
\cos \vartheta \leq 1 \]
{}From \qquad $(\cos \vartheta )_{min} \leq 1 $ \qquad the lower limit of
$E_{q_{out}}$ is obtained as
\[  E_{q_{out}}^{min}=\frac{Q^{2}_{cut}}{4E_{cm}^{eP}\sqrt{x}} \]
and finally from
\qquad $ E_{q_{out}}^{min} \leq E_{q_{out}}^{max} $\qquad the region of
$x$ integration is fixed as
\[ \frac{Q^{2}_{cut}+(\ml +\mxi )^2}{4(E_{cm}^{eP})^2}  \leq x \leq 1 .\]
\section*{Acknowledgements}
The authors would like to thank F.~Franke for many helpful discussions.
T.W. would also like to thank X.Tata for many fruitfull discussions and
carefully reading of the manusript.
T.W. is supported by the {\em Deutsche Forschungsgemeinschaft (DFG)}.
All numerical calculations were performed at the
{\em Rechenzentrum der Universit\"at W\"urzburg}. This research was supported,
in part, by the U.S. Dept. of Energy Grant No. DE-FG-03-94ER40833.
 \newpage
\section*{Table Captions}
Table 1: Signfactor $\theta_i$,
with $\varepsilon_A = \mbox{ sign}(M \sin \beta +\mu \cos \beta )$
and $\varepsilon_B = \mbox{ sign}(M \cos \beta +\mu \sin \beta )$.\\
Table 2: Gauge boson contribution to Feynman graphs 1, 3 and 5 for the
basic subprocesses.\\
Table 3: Neutralino and chargino states in three different mixing
scenarios (A), (B) and (C). Shown are the masses as well as the
coefficients $N_{ik}$ from eq. (\ref{nffkop}) and $V_{ij}$ from eq.
(\ref{cffkop}) of the decomposition into the weak eigenstates.
\section*{Figure Captions}
Fig. 1: Feynman diagrams for the basic subprocess $eq
\rightarrow \tilde l \tilde \chi_{i}^{0(\pm)} q$. The gauge bosons
exchanged in graphs 1, 3, 5 are denoted by $X$ (see table 2).
\newline
Fig. 2: Cross sections in scenario (A1) for the processes $ep
\rightarrow \tilde e \tilde \chi_{i}^{0} X$, with dashed line for
$i$=1, dotted line for $i$=2, dash-dotted for $i$=3, dash-dot-dot for
$i$=4 and solid line for $ep
\rightarrow \tilde e \tilde q X$.\newline
Fig. 3: The same as fig. 2 for scenario (A2).\newline
Fig. 4: The same as fig. 2 for scenario (C1).\newline
Fig. 5: The same as fig. 2 for scenario (C2).\newline
Fig. 6: Cross sections in scenario (A1) for the processes $ep
\rightarrow \tilde e \tilde \chi_{i}^{\pm} X$, with dashed line for the
production of $\tilde \chi_{1}^{-}$,
dotted line for $\tilde \chi_{2}^{-}$,
dash-dotted for $\tilde \chi_{1}^{+}$, dash-dot-dot for
$\tilde \chi_{2}^{+}$ and solid line for $ep
\rightarrow \tilde e \tilde q X$.\newline
Fig. 7: The same as fig. 6 for scenario (A2).\newline
Fig. 8: Cross sections in scenario (A1) for the processes $ep
\rightarrow \tilde \nu \tilde \chi_{i}^{0} X$, with dashed line for
$i$=1, dotted line for $i$=2, dash-dotted for $i$=3, dash-dot-dot for
$i$=4 and solid line for $ep
\rightarrow \tilde \nu \tilde q X$.\newline
Fig. 9: The same as fig. 8 for scenario (A2).\newline
Fig. 10: The same as fig. 8 for scenario (A').\newline
Fig. 11: The same as fig. 8 for scenario (B1).\newline
Fig. 12: The same as fig. 8 for scenario (B2).\newline
Fig. 13: The same as fig. 8 for scenario (C1).\newline
Fig. 14: The same as fig. 8 for scenario (C2).\newline
Fig. 15: Cross sections in scenario (A1) for the processes $ep
\rightarrow \tilde \nu \tilde \chi_{i}^{-} X$, with dashed line for
$i$=1, dotted line for $i$=2 and solid line for $ep
\rightarrow \tilde \nu \tilde q X$.\newline
Fig. 16: The same as fig. 15 for scenario (A'), with the
masses of the charginos $\tilde \chi_{1}^{-}$ and $\tilde
\chi_{2}^{-}$ included.\newline
Fig. 17: The same as fig. 15 for scenario (B1).\newline
Fig. 18: The same as fig. 16 for scenario (B').\newline
Fig. 19: The same as fig. 15 for scenario (C1).\newline
Fig. 20: The same as fig. 16 for scenario (C').
\newpage
\begin{center}
\begin{tabular}{|c|c|c|}
\hline
 & $\tan \beta >1 $ & $\tan \beta <1$ \\
\hline
$ \theta_1 $ & 1 & $\varepsilon_B $ \\
$ \theta_2 $ & $\varepsilon_B $ & 1  \\
$ \theta_3 $ & $\varepsilon_A $ & 1  \\
$ \theta_4 $ & 1 & $\varepsilon_A $ \\
\hline
\end{tabular} \\[0.5cm] Table 1
\end{center} \newpage
\begin{center}
\begin{tabular}{|c|c|c|c|c|c|}
\hline
 & (a) $ eq \rightarrow \tilde e \tilde \chi_{i}^{0} q $ &
 (b) $eq \rightarrow \tilde e \tilde \chi_{i}^{+} q' $ &
 (c) $eq \rightarrow \tilde e \tilde \chi_{i}^{-} q' $ &
 (d) $eq \rightarrow \tilde \nu \tilde \chi_{i}^{0} q' $ &
 (e) $eq \rightarrow \tilde \nu \tilde \chi_{i}^{-} q $
\\ \hline
$ {\cal M}_1 $ & $X=\gamma,\, Z$ & -- & $X=W$ & $X=W$ & $X=Z$ \\ \hline
$ {\cal M}_3 $ & $X=\gamma,\, Z$ & $X=W$ & -- & $X=W$ & $X=\gamma,\, Z$
\\ \hline
$ {\cal M}_5 $ & $X=Z$ & $X=W$ & $X=W$ & $X=W$ & $X=\gamma,\, Z$ \\
\hline
\end{tabular}\\[0.5cm] Table 2
\end{center} \newpage
\begin{tabular}{|c||c|c|c|}
\hline
 & A & B & C \\ \hline \hline
$\tan \beta$ &2&2&2 \\ \hline
$\mu$ & -219 GeV & 119 GeV & -44 GeV \\ \hline
M & 73 GeV & 169 GeV & 219 GeV \\ \hline
$\tilde \chi_{1}^{0}$ & $m=40$ GeV, $\eta=+1$& $m=40$ GeV, $\eta=+1$&
 $m=40$ GeV, $\eta=+1$\\
 & (-0.95,+0.30,+0.08,+0.08) & (-0.35,+0.63,-0.62,-0.33) &
(-0.06,+0.13,-0.18,+0.97) \\ \hline
$\tilde \chi_{2}^{0}$ & $m=88$ GeV, $\eta=+1$& $m=106$ GeV, $\eta=+1$&
$m=74$ GeV, $\eta=-1$\\
 & (-0.32,-0.89,-0.18,-0.27) & (-0.91,-0.06,+0.39,+0.14) &
(+0.07,-0.33,+0.92,+0.22) \\ \hline
$\tilde \chi_{3}^{0}$ & $m=225$ GeV, $\eta=+1$& $m=122$ GeV,
$\eta=-1$& $m=118$ GeV, $\eta=+1$\\
 & (+0.02,+0.20,+0.35,-0.92) & (+0.02,-0.12,+0.35,-0.93) &
(+0.92,-0.32,-0.20,+0.06) \\ \hline
$\tilde \chi_{4}^{0}$ & $m=244$ GeV, $\eta=-1$& $m=230$ GeV,
$\eta=+1$& $m=243$ GeV, $\eta=+1$\\
 & (+0.01,-0.27,+0.92,+0.29) & (+0.22,+0.77,+0.59,+0.13) &
(+0.37,+0.88,+0.29,-0.04) \\ \hline
$\tilde \chi_{1}^{+}$ & $m=87$ GeV, $\eta=+1$& $m=66$ GeV, $\eta=-1$& $
m=61$ GeV, $\eta=+1$\\
 & (+0.99,+0.10) & (+0.65,-0.76) & (+0.39,-0.92) \\ \hline
$\tilde \chi_{2}^{+}$ & $m=241$ GeV, $\eta=+1$& $m=225$ GeV,
$\eta=+1$& $m=242$ GeV, $\eta=+1$\\
 & (-0.10,+0.99) & (+0.76,+0.65) & (+0.92,+0.39) \\ \hline
\end{tabular} \\[0.5cm] \begin{center} Table 3 \end{center} 
\end{document}